\documentclass[aps, pre, amsmath, 11pt, showkeys]{revtex4-2}

\usepackage{graphicx}
\usepackage{mathtools}
\usepackage{amssymb,amsmath}
\usepackage{booktabs}
\usepackage{multirow}
\usepackage{tabularx}

\usepackage[hidelinks]{hyperref}

\begin{document}

\title{Early Warnings for Multistage Transitions in Dynamics on Networks}

\author{Neil G. MacLaren$^1$}
\author{Prosenjit Kundu$^{1,\dagger}$}
\author{Naoki Masuda$^{1,2}$}
\email{Correspondence: naokimas@buffalo.edu}

\affiliation{$^1$Department of Mathematics, State University of New York at Buffalo, NY 14260-2900, USA}
\affiliation{$^2$Computational and Data-Enabled Science and Engineering Program, State University of New York at Buffalo, Buffalo, NY 14260-5030, USA}
\affiliation{$^{\dagger}$Present address: Dhirubhai Ambani Institute of Information and Communication Technology, Gandhinagar, Gujarat 382007, India.}
\date{\today}

\begin{abstract}
  Successfully anticipating sudden major changes in complex systems is a practical concern. Such complex systems often form a heterogeneous network, which may show multistage transitions in which some nodes experience a regime shift earlier than others as an environment gradually changes. Here we investigate early warning signals for networked systems undergoing a multistage transition. We found that knowledge of both the ongoing multistage transition and network structure enables us to calculate effective early warning signals for multistage transitions. Furthermore, we found that small subsets of nodes could anticipate transitions as well as or even better than using all the nodes. Even if we fix the network and dynamical system, no single best subset of nodes provides good early warning signals, and a good choice of sentinel nodes depends on the tipping direction and the current stage of the dynamics within a multistage transition, which we systematically characterize.
\end{abstract}


\maketitle

\raggedbottom

\section{Introduction}
\label{sec:intro}

A characterization of complex systems is dependence among components, which often leads to surprising, nonlinear behavior. One important nonlinear phenomenon is that of a tipping point: a transition in which stable aspects of the system suddenly shift to a drastically altered state when the system's environment changes by a small amount; recovery from the altered state is typically difficult. Tipping points have been described in, for example, the switch from clear to turbid water in lake ecosystems \cite{scheffer1993}, changes in fish community composition \cite{carpenter2011}, alterations in global climate regimes \cite{reid2016}, and in the progression of disease \cite{venegas2005,foo2017}. This shared feature of such disparate systems can be described mathematically by bifurcations, and several early warning signals---statistical indications that a bifurcation point is nearby---have been developed that attempt to anticipate such transitions.
These early warning signals rely on a process called critical slowing down: systems recover from perturbations more slowly near a bifurcation point \cite{scheffer2009}. Critical slowing down results in predictable signatures in time series data, including increasing variance and autocorrelation, and it is these signatures that are used to construct early warning signals. 
Early warning signals based on the critical slowing down phenomenon have been validated in several model systems \cite{carpenter2011,dai2012}, and their practical utility has been demonstrated in, e.g., predicting electrical grid failures \cite{hines2011} and reversing cyanobacterial blooms \cite{pace2017}.

Many systems showing tipping points can be modeled by a network in which a node represents a dynamical system and different dynamical systems interact through the edges of the network \cite{gross2008}. Studying tipping points in such systems is an integral part of studying network robustness and resiliency \cite{liu2022}. An example with applications in conservation ecology is the anticipation of a breakdown in mutualistic species networks \cite{lever2020,dakos2018,aparicio2021}. In such models, species populations are typically represented by stochastic differential equations interacting through a bipartite network of plants and pollinators \cite{lever2014,dakos2014} or a unipartite projection focusing on only plants or pollinators \cite{lever2020,gao2016}. Early warning signals can then predict major adjustments in species composition \cite{lever2020} or population collapse \cite{aparicio2021}. Similarly, exploiting information on interactions between weather patterns in different regions may improve the forecasting of climate tipping points \cite{wunderling2020}. 

In fact, the inherent heterogeneity in networked systems may make tipping points more complex. Specifically, multistage transitions, in which not all components transition to an alternate state at the same parameter values, may be the rule rather than the exception in networks with certain features \cite{kundu2022,kundu2022b}. Multistage transitions have been documented in studies of mutualistic species dynamics \cite{lever2020,aparicio2021} and climate systems \cite{wunderling2020}, and are consistent with evidence from human commensal bacteria \cite{lahti2014} and social upheaval \cite{brummitt2015}. The ability to anticipate multistage transitions would thus have applications in many fields. 

A variety of methods have been proposed to provide early warning of tipping points on networks.
Examples include aggregations of univariate (i.e., single-node) early warning signals and explicitly multivariate methods such as measures derived from a principal component analysis (PCA) of state variables \cite{weinans2021,liu2022}. However, most of the available early warning signals for networks treat the network as a united entity and do not exploit the fact that a network is composed of subsystems that may show different dynamics and provide different early warning signals. There are some notable exceptions.
First, Chen et al. used cross-correlations to identify clusters of nodes that were more sensitive to an approaching bifurcation than the network as a whole~\cite{chen2012}. Although Chen et al. exploited network heterogeneity for constructing early warning signals, they did not consider multistage transitions.
Second, Lever et al. developed PCA methods to predict the direction and magnitude of change for each node's state after a bifurcation~\cite{lever2020}. Lever et al. noted parameter ranges for their model in which multistage transitions were possible and that the early warning signal they proposed tended to correctly anticipate the first transition. However, Lever et al. noted that their method was less reliable for describing further nodes' transitions---the multistage transition.
Third, Aparicio et al. used network control theory---rather than system dynamics---to identify nodes that would be capable of providing a reliable early warning signal~\cite{aparicio2021}. They also identified parameter values that caused multistage transitions in their model and also found that their method underperformed in those regions. In contrast to Lever et al.'s method, Aparicio et al.'s method tended to miss early transitions of nodes but correctly predicted the final collapse. Based on the ubiquitousness of multistage transitions in networks, discussed above, there is a need for early warning signals that can provide alerts for each of the major tipping points within a multistage transition that a networked system may experience.

In the present study, we build on key points from these three studies---namely that (1) some nodes may be more informative about impending transitions than others and (2) information may be available in the network structure or dynamics with which to anticipate multistage transitions---to investigate early warning signals for multistage transitions in tipping dynamics on networks. We find that traditional early warning signals are in fact able to provide early warning in a network undergoing a multistage transition. Using knowledge of the network allows us to choose ``sentinel'' nodes, i.e., node sets that can provide early warning more efficiently than using all nodes in terms of the number of nodes we must observe. Furthermore, it is often the case that such early warning signals even improve in accuracy.

\section{Methods}
\label{sec:methods}

\subsection{Model}
\label{sub:model}

Consider an undirected and unweighted network of $N$ nodes and denote its adjacency matrix by $A = (a_{ij})$ with $a_{ii} = 0$ and $a_{ij} = a_{ji} \in \{0, 1\}~\forall~i, j \in \{1, \ldots, N \}$. We simulate the stochastic dynamics of a coupled double-well model on networks given by
\begin{equation}
  \label{eq:doublewell}
  \frac{dx_i}{dt} = -(x_i - r_1)(x_i - r_2)(x_i - r_3) + D\sum_{j=1}^N a_{ij}x_j + s\xi_i,
\end{equation}
where $x_i$ is the state of node $i$; $r_1$, $r_2$, and $r_3$ are parameters that control the location of the equilibria and satisfy $r_1 < r_2 < r_3$; $D~(\ge 0)$ is the coupling strength; and $s \xi_i$ is a Gaussian noise process with standard deviation $s$. The first term is the derivative of a fourth-order polynomial representing a double-well potential. In the uncoupled and noiseless case, it produces lower and upper stable equilibria at $x_i=r_1$ and $x_i =r_3$, respectively, and an unstable equilibrium at $x_i = r_2$, and it also creates hysteresis.
Unless we state otherwise, we set $(r_1, r_2, r_3) = (1, 4, 7)$.
The coupling term $D\sum_{j=1}^N a_{ij}x_j$ shifts $x_i$ at the stable equilibria out of $x_i = r_1 = 1$ or $x_i = r_3 = 7$.
In addition, the noise term $s \xi_i$ lets $x_i$ jitter around the stable equilibria obtained in the absence of noise.
We therefore consider that nodes with $x_i < 2.268$ are in the lower state and $x_i > 2.268$ are in the upper state. We selected this threshold value for $x_i$ because the cubic term in Eq.~\eqref{eq:doublewell} has an inflection point at $x_i \approx 2.268$ in the absence of the coupling term, demarcating a basin of attraction for the lower stable point at $x_i = 1$. We numerically verified that we can reliably classify $x_i$ into the lower and upper stable equilibria with these threshold values even in the presence of the coupling term (see Figure~\ref{fig:demo}).
Equation~\eqref{eq:doublewell} represents dynamics of species abundance \cite{lever2020} or climates in interconnected regions \cite{wunderling2020}. We primarily consider $D$ as a bifurcation parameter. A possible mechanism underlying variation in $D$ is the volume of moisture moving from one climate basin to another \cite{wunderling2020}.

For applications such as species loss in population ecology, one is interested in beginning with the upper state, which corresponds to the situation in which all the species are abundant, and gradually varying a parameter value to anticipate transitions of various nodes to their lower states \cite{scheffer2009}. For example, a transition to the lower state could correspond to the collapse of a species' population. To validate the relevance of multistage transitions and early warning signals in this scenario, we consider an extension of Eq.~\eqref{eq:doublewell} given by
\begin{equation}
  \label{eq:alt}
  \frac{dx_i}{dt} = -(x_i - r_1)(x_i - r_2)(x_i - r_3) + D\sum_{j=1}^N a_{ij}x_j + u + s\xi_i.
\end{equation}
Variable $u$ is a stressor that directly and uniformly influences all nodes. An increase in $u$ represents, for example, increased global mean temperature \cite{wunderling2020} or degradation of the local environment causing increased mortality for all species \cite{lever2020}. 
With Eq.~\eqref{eq:alt}, we hold either $D$ or $u$ constant and vary the other as the bifurcation parameter.

\subsection{Numerical Simulations}
\label{sub:sims}

Unless we state otherwise, we used $D$ as the bifurcation parameter and began simulations with all nodes in the lower state. For the given network and the value of $D$, we started the dynamics from the initial condition $x_1 = \cdots = x_N = 1$. For any given value of $D$, we integrated Eq.~\eqref{eq:doublewell} using the Euler-Maruyama method with time step $\Delta t = 0.01$ for 50 time units (TU) to allow $\{x_1, \ldots, x_N \}$ to relax to an equilibrium. In fact, allowing 50 TU was sufficient except in rare cases in which some nodes changed their macroscopic state (i.e., lower versus upper state) after 50 TU due to dynamical noise. We then continued simulating the dynamics for a further 25 TU to take samples from $\{ x_1(t), \ldots, x_N(t)\}$ for calculating early warning signals. We used $s = 0.05$ except where noted.

To determine whether or not early warning signals increase prior to transitions of various nodes from their lower state to upper state, we conducted sequences of the above simulations for a given network and set of parameters. Each sequence began with $D = 0.01$. After we simulated the dynamics for 75 TU in total and calculated early warning signals, we increased $D$ by 0.005, reset $x_i~\forall~i$ to the initial condition, ran the simulation with the new value of $D$, and calculated early warning signals from the new $x_i(t)$. We continued this procedure (i.e., increasing $D$ by 0.005 and running a new simulation) until at least 90\% of nodes reached the upper state at equilibrium.

In simulations with $D$ as the bifurcation parameter but with the nodes beginning in the upper state, we set $x_i = 7~\forall~i$ and $u = -15$. In this case,
we consider that nodes with $x_i < 5.732$ are in the lower state and $x_i > 5.732$ are in the upper state; note that
Eq.~\eqref{eq:doublewell} has a second inflection point at $x_i \approx 5.732$ in the absence of the coupling term. We initially set $D = 1$ and decreased $D$ by 0.005 in each simulation, continuing until $> 90$\% of nodes transitioned to the lower state at equilibrium. All other parameters were the same regardless of whether we began simulations with the nodes at the upper or lower state.

This simulation method attempts to ensure that we always study the system at equilibrium and has been used in previous studies of tipping points on networks (e.g., \cite{wunderling2020}). De-trending or other preprocessing of data from the simulations is therefore not needed: by the time we take data from each simulation, the system is stationary by design (c.f. \cite{dakos2012} for a different simulation method, for which de-trending is required). 

\subsection{Early Warning Signals}
\label{sub:earlywarnings}

At each value of $D$, we calculated the following early warning signals \cite{weinans2021, dakos2012} from $M=250$ equally spaced samples of $\{x_1(t), \ldots, x_N(t)\}$ with $t \in (50, 75]$, i.e., with $t~\in~\{50.1, 50.2, \ldots 75.0\}$:
\begin{itemize}
\item The dominant eigenvalue $\lambda_{\rm max}$ of the covariance matrix, of which the $(i, j)$ entry is the covariance of $\{x_i(50.1), x_i(50.2), \ldots, x_i(75)\}$ and $\{x_j(50.1), x_j(50.2), \ldots, x_j(75)\}$.
\item The standard deviation of each $x_i(t)$ estimated from the $M$ samples.
\item The lag-1 autocorrelation of each $x_i(t)$, defined as $\frac{\sum_{m=1}^{M-1} (x_{i,m} - \overline{x}_i)(x_{i,m+1} - \overline{x}_i)}{\sum_{m=1}^M (x_m - \overline{x}_i)^2}$, where $x_{i,m} \equiv x_i(50 + 0.1 m)$ and $\overline{x}_i = \sum_{m=1}^M x_{i, m} / M$.
\end{itemize}
To define an early warning signal for a given node set, we used both the maximum and the mean of the standard deviation and lag-1 autocorrelation in addition to $\lambda_{\rm max}$ calculated from the node set of interest. Therefore, we examine five different early warning signals for a given set of nodes (see section~\ref{sub:node-set} for the node sets).

We quantify the extent to which an early warning signal anticipates a bifurcation with the Kendall rank correlation, $\tau$, between $D$ before the bifurcation occurs and the early warning signal  \cite{dakos2008}. The reasoning behind using Kendall's $\tau$ as a performance metric is as follows. Consider a range of $D$ in which no nodes change state at equilibrium except at the final value of $D$. We refer to a range of $D$ in which the number of nodes in the lower/upper state is constant as a stable range. Given our simulation protocol, $D$ is linearly increasing in a stable range. If an early warning signal tends to increase as $D$ increases towards the bifurcation point, indicating critical slowing down, then the early warning signal is considered to be useful in anticipating the bifurcation, and $\tau$ tends to be large. However, in the network dynamics that we are considering, there are potentially many values of $D$ at which some nodes switch from the lower to the upper state. Therefore, we correlate $D$ with a given early warning signal to obtain $\tau$ only within stable ranges of $D$ having at least 15 unique values of $D$. We report the $\tau$ value averaged over all such stable ranges. For example, if there is no node transitioning from its lower state to the upper state for $D \in \{0.01, 0.015, \ldots, 0.5\}$, $D \in \{0.505, 0.51, 0.515\}$, and $D \in \{0.52, 0.525, \ldots, 1\}$, some nodes transit from the lower to the upper state at $D = 0.505, 0.52$, and $1.005$, and the transition at $D=1.005$ makes the fraction of the nodes in the upper state exceed $0.9$, then we calculated $\tau$ for the first and third ranges of $D$ and took the average of the two $\tau$ values.

\subsection{Node Sets}
\label{sub:node-set}

We defined the following nine node sets for calculating the early warning signals:
\begin{itemize}
\item ``All'' refers to the set of all nodes.
\item ``Lower State'' refers to the set of all nodes in the lower state at $t = 50$ TU.
\item ``Upper State'' refers to the set of all nodes in the upper state at $t = 50$ TU. If there are no nodes in the upper state, this node set is empty and early warning signals for this node set are undefined.
\item ``High Input'' refers to the $n$ nodes that are largest in terms of $R_i = \sum_{j=1}^N a_{ij}\overline{x}_j$, where $i$ is the index of an available node in the sense that it is still in its original macro state. For example, a lower-state node is an available node if nodes are initially in the lower state in a simulation. Note that such a node is available to transition to the upper state as $D$ increases. We remind that $\overline{x}_j$ is the mean of $x_j$ calculated over the $M$ samples. We define the High Input node set based on the idea that a lower-state node with many neighbors or with neighbors in the upper state is more likely to transition from the lower to the upper state earlier than other nodes.
\item ``Low Input'' refers to the $n$ nodes that are the smallest in terms of $R_i$. As for High Input, we require that the $i$th node is in its original macro state. The Low Input node set reflects the observation that, if the nodes are initially in the upper state, then the node with the smallest contribution from the coupling term, i.e., those with smallest $R_i$, would be the first to transition to the lower state as $D$ decreases.
\item ``Lower Half'' refers to the set of lower-state nodes below the median in terms of $R_i$; we do not use this node set when all the nodes are initially in the upper state in the simulation. If the nodes begin in the lower state, Lower Half nodes are the farthest from a bifurcation as one gradually increases $D$.
\item ``Random'' refers to the set of $n$ nodes selected uniformly at random.
\item ``Large Correlation'' nodes are the top $n$ nodes in terms of $R'_i = \sum_{j=1; j\neq i}^N \text{cor}(x_i, x_j)\overline{x}_j$, where the $i$th node is a lower-state node, and $\text{cor}(x_i, x_j)$ is the Pearson correlation coefficient between $x_i$ and $x_j$ calculated over the $M$ samples. This is an alternative for High Input when we do not have access to the network structure, i.e., the adjacency matrix.
\item ``Large Standard Deviation (Large SD)'' nodes are the $n$ nodes with the largest standard deviation of $x_i$ over the $M$ samples. A node tends to have a larger standard deviation when it receives a larger input from the coupling term. Thus, the Large SD node set is also an alternative for High Input when we do not have information about the network structure.
\end{itemize}
The All node set corresponds to established early warning signal methods and is the most costly in terms of sampling effort. The High Input, Low Input, Random, Large Correlation, and Large SD node sets require a limited number of nodes, which we set $n = 5$, and are therefore the least costly. 
The other node sets are variable in terms of the number of nodes. However, with the exception of the first stable range, the number of nodes used is typically much larger than $n$ and much smaller than $N$ across a wide range of $D$. All, Lower State, Upper State, Random, Large Correlation, and Large SD do not use the information on the network structure, whereas High Input, Low Input, and Lower Half do. Random, Large Correlation, and Large SD are most economic in the sense that it only uses $n$ nodes and does not require the network structure.
We updated node set membership each time we change the value of $D$.

\subsection{Networks}
\label{sub:networks}

We conducted simulations on 6 model networks and 17 empirical networks; see the Supplementary Information (SI) for details of the networks. We chose networks having the order of 100 nodes, similar in size to many empirical networks and small enough to be computationally feasible for our simulations. We chose model networks with a range of degree heterogeneities and with and without a planted community structure, including networks that show a multistage transition to different extents~\cite{kundu2022b}. Empirical networks may have a variety of features difficult to capture with model networks and thus present hidden challenges to our methods.
An example of our empirical networks is a dolphin social network \cite{lusseau2003}. In this network, the nodes are individual dolphins and two nodes are adjacent if individuals $i$ and $j$ were observed together more often than expected by chance. On such a network, $x_i$ represents, for example, a behavioral state or possession of particular information.

\subsection{Robustness Analysis}
\label{sub:robustness}

We tested several variations of our methods to examine robustness under different scenarios. First, 
to test the robustness of these results with respect to the network structure, we conducted simulations on the 23 networks explained in Section~\ref{sub:networks}. Ten of the 23 networks had at least two stable ranges, showing clear multistage transitions. We selected these ten networks for further analysis.

Consider an early warning signal. On each of the ten selected networks, we calculated $\tau$ between the early warning signal and $D$ for each stable range of $D$. We then averaged $\tau$ over the stable ranges of $D$. We calculated such an averaged $\tau$ value 50 times, restarting simulations with a new random seed each time, for each of the three node sets (i.e., All, Lower State, and High Input) and each network. Finally, we estimated a linear mixed effects model to predict the averaged $\tau$ value based on three levels of a node-set fixed effect variable (i.e., All as the reference, Lower State, and High Input) with a random effect for network. We estimated the linear mixed effects model in this manner for each of the five early warning signals.

Second, we varied several simulation parameters on two arbitrarily selected networks. The adjusted parameters were the noise intensity ($s \in \{0.01, 0.1, 0.5\}$), the number of samples taken from each $x_i(t)$ when calculating early warning signals ($M \in \{25, 50, 150\}$), the double-well model parameters ($(r_1, r_2, r_3) \in \{(1, 3, 5),~(1, 2.5, 7),\allowbreak~(1, 5.5, 7)\}$), and the duration $T$ of the simulation before we start to sample $\{x_1(t), \ldots, x_N(t)\}$ to calculate the early warning signals at each value of $D$ ($T \in \{25, 75, 100 \}$).

Third, we altered the model itself, examining transitions from the upper to the lower state using Eq.~\eqref{eq:alt}. 

\subsection{Software}
\label{sub:software}

We conducted all simulations and analyses in R (v4.2); dependencies include the ``igraph'' package (v1.3) for network analysis \cite{csardi2006}, the ``nlme'' package (v3.1) for mixed effects statistical models \cite{pinheiro2021}, and the ``parallel'' package (v4.2) \cite{R} for parallel processing. Empirical networks were drawn from the ``networkdata'' package \cite{schoch2022}. Code and data to reproduce these analyses are available at \href{https://github.com/ngmaclaren/doublewells}{https://github.com/ngmaclaren/doublewells}.

\section{Results}
\label{sec:results}

\subsection{Multistage Transitions and Performance of Early Warning Signals Based on Different Node Sets}

\begin{figure}
  \centering
  \includegraphics[width = .7\textwidth]{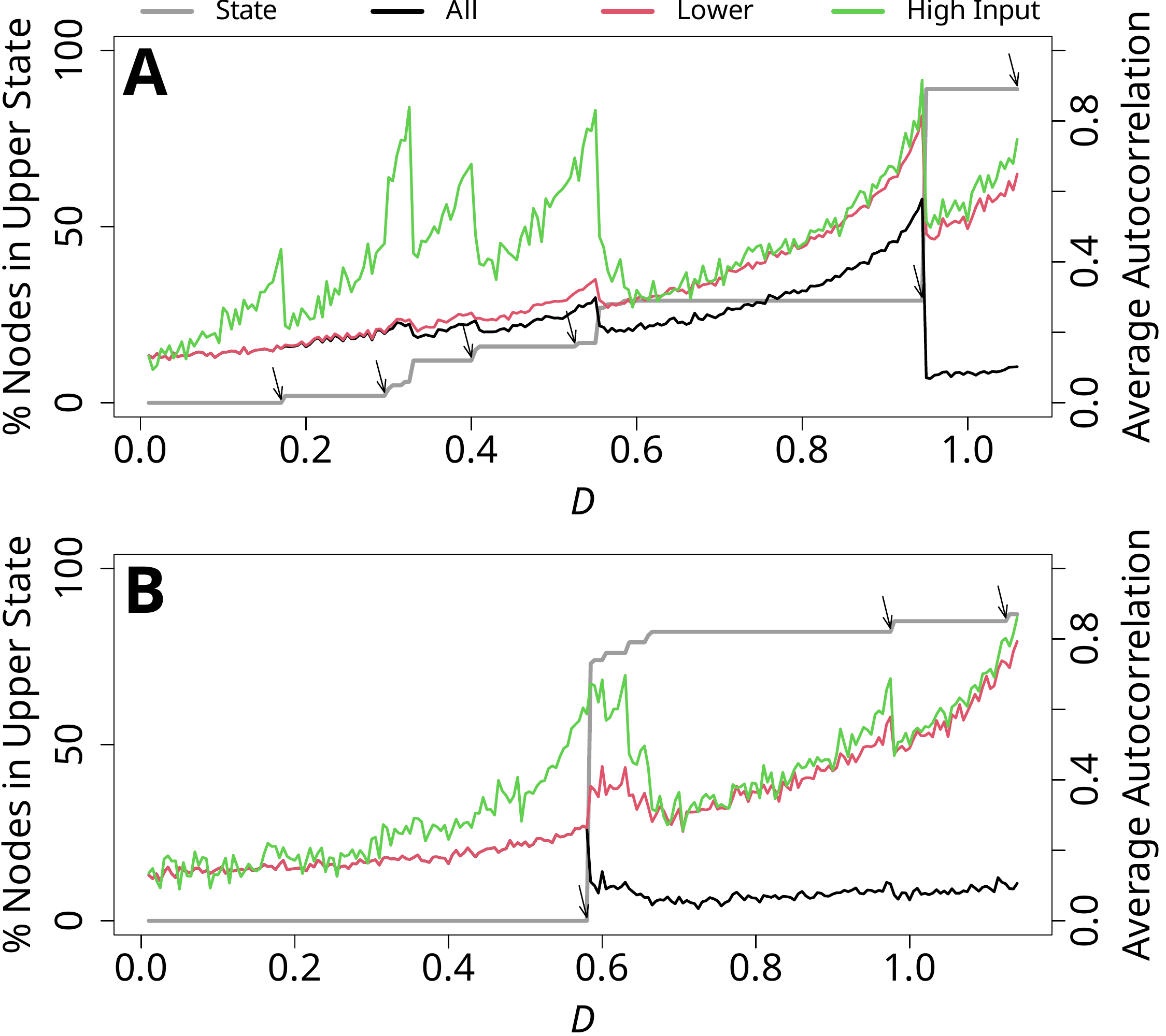}
  \caption{Multistage transitions when the nodes are initially in the lower state. We show the number of nodes in the upper state at equilibrium (gray), and the average lag-1 autocorrelation of $x_{i,t}$ calculated for all nodes (black), the nodes in the lower state (red), and the low-input nodes (green). The arrows mark transitions of some nodes at the ends of stable ranges. (A) A network with 100 nodes and a power-law degree distribution; (B) Dolphin social network.}
  \label{fig:examples}
\end{figure}

Let us first consider a network with $100$ nodes and a power-law degree distribution generated by a configuration model, which we call the power-law network.
We show by the gray line in Figure~\ref{fig:examples}A the proportion of nodes in the lower state in the equilibrium as a function of the coupling strength between nodes, $D$.
The figure shows that more nodes tend to be in the upper state in the equilibrium when $D$ is larger. Additionally, there are ranges of $D$ in which relatively large changes in $D$ do not induce transition of any node from the lower to the upper state at equilibrium. In other ranges of $D$, small changes in $D$ trigger transitions of some nodes between macro states. In this manner, the noisy double-well model on this network shows a multistage transition.
We also find a multistage transition when we use Eq.~\eqref{eq:alt} and vary $u$ instead of $D$ as the bifurcation parameter (Fig.~\ref{fig:u1}).

Early warning signals appear to be sensitive to changes in $D$. Figure \ref{fig:examples}A also shows a typical early warning signal, i.e., the lag-1 autocorrelation of $x_i(t)$, averaged over three different node sets. The first, ``All'' (black), corresponds to traditional early warning signals and refers to the set of all nodes. Within the stable ranges of $D$, the early warning signal value tends to increase as $D$ increases. However, different nodes may be differently informative as to an impending transition. Both observed dynamics \cite{chen2012,lever2020} and knowledge of network structure \cite{aparicio2021} may improve the accuracy of early warning signals or their efficiency in terms of the amount of observed signals necessary for the calculation. In fact, it may be more efficient to monitor nodes that are most likely to transition to an alternate state with a perturbation of a control parameter.

To show that monitoring sentinel node sets can be effective, Figure \ref{fig:examples}A also displays the early warning signal calculated for the set of nodes in the lower state at $t = 50$ TU (``Lower State'', red) and the set of five nodes most likely to transition from the lower to upper state (``High Input'', green). These latter nodes have many neighbors, are connected to nodes that have transitioned to the upper state, or both; they have the highest value of $R_i \equiv \sum_{j=1}^N a_{ij}\overline{x}_j$ by definition. Figure \ref{fig:examples}A shows that the sensitivity of the average autocorrelation to the increase in $D$ towards the end of a stable range varies depending on the node set and the value of $D$. For example, there is a major sudden increase in the number of nodes in the upper state at equilibrium at $D = 0.95$. This transition is associated with, looking from left to right, a marked increase in the average autocorrelation of the nodes in each of the node sets at $D$ just below 0.95 and a decrease in the average autocorrelation at $D = 0.95$. A similar tendency is present around the transitions of smaller batches of nodes at, for example, $D = 0.175$, 0.33, and 0.55. Changes in the average autocorrelation of the High Input nodes tend to be larger in absolute value than for the Lower State and All node sets, particularly at smaller values of $D$, but the overall range is similar in this network.

Figure~\ref{fig:examples}B shows that the double-well model on a dolphin social network \cite{lusseau2003} also exhibits a multistage transition.
See Fig.~\ref{fig:u1} for similar results when $u$ is the bifurcation parameter.
Compared to the case of the power-law network, the dolphin network allows larger stable ranges of $D$, and the ranges of $D$ in which small changes in $D$ induce a transition of a notable fraction of nodes from the lower to the upper state are narrower. Similar to Fig.~\ref{fig:examples}A, the autocorrelation tends to reliably increase in each stable range of $D$ as we increase $D$ towards the value at which some nodes transit from the lower to the upper state.
In addition, the average autocorrelation based on the Lower State and High Input node sets apparently better signals such transitions than that based on all nodes in
the sense that the average autocorrelation increases more drastically as $D$ increases towards the bifurcation.

\begin{figure}
  \centering
  \includegraphics[width = \textwidth]{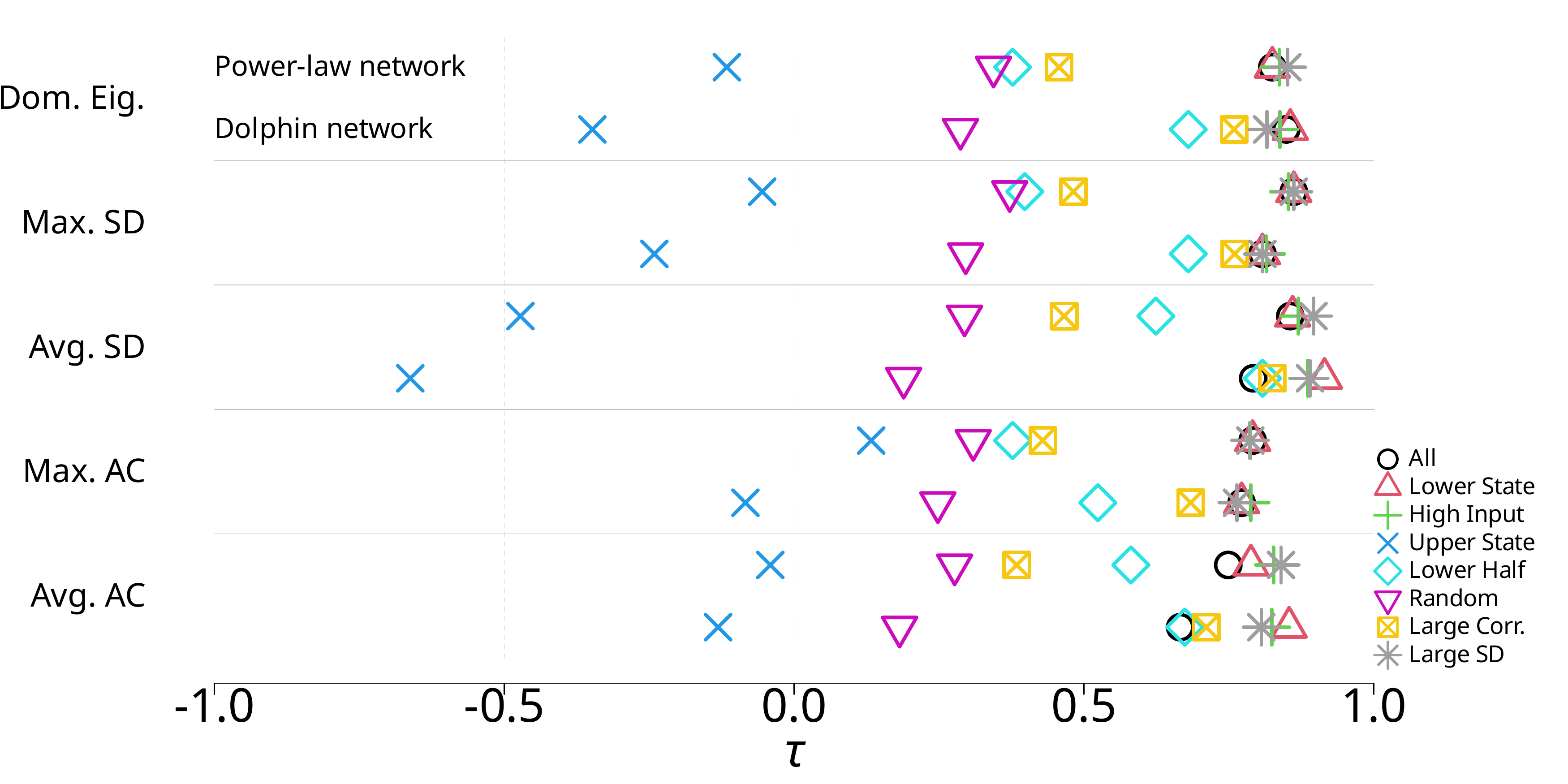}
  \caption{Kendall correlations ($\tau$) between each of the five early warning signals and the coupling strength, $D$, for different sets of nodes. See main text for details of node set membership. Dom. Eig: dominant eigenvalue of the covariance matrix of all nodes in the node set, Max. SD, Avg. SD: maximum and average standard deviation of $x_i$, Max. AC and Avg AC: maximum and average autocorrelation of $x_i$, Large Corr.: the Large Correlation node set.}
  \label{fig:examples-taus}
\end{figure}

To quantify the performance of the average autocorrelation and other early warning signals, we computed the Kendall's $\tau$ for each of the two networks used in Fig.~\ref{fig:examples} and for each of the five early warning signals calculated for each node set. We show the results in Fig.~\ref{fig:examples-taus}, which indicates that $\tau$ is high (i.e., $> 0.65$) across both networks and all five early warning signals and for All (circles), Lower State (triangles), and High Input (pluses) node sets. The $\tau$ values for each early warning signal in both networks are similar between Lower State and High Input, and they are higher than for All in a majority of cases. 
In addition to having a high average $\tau$ value, the High Input node set has $\tau > 0.7$ for each major transition in both networks (see SI section \ref{sec:eachtransition} and Fig.~\ref{fig:each} for details). If we calculate the average autocorrelation for the nodes that actually changed state at each major transition, we of course find that the $\tau$ value for this retroactively identified node set is high. However, the High Input node set has almost the same performance, in terms of $\tau$ at each transition, as the nodes that actually changed state (Fig.~\ref{fig:each}). 
Furthermore, by definition, early warning signals calculated with the Lower State and High Input node sets are more cost-efficient than those calculated with all nodes because the former use only a fraction of nodes.
However, our typical simulations use samples of $x_i$ at all $M$ time points for both assigning nodes to node sets and calculating early warning signals. We performed additional simulations, described in section \ref{sec:sampling}, which only used the samples at the first ten time points to determine node set membership. We then monitored the node set members for the full $M$ samples including the first ten samples for calculating early warning signals. Our results are robust to this decision, as we show in Fig.~\ref{fig:sampling}.
Finally, the High Input node set performs well even when we consider all node transitions, not just those occurring after a stable range (Fig.~\ref{fig:errors}).

Although the Lower State node set is both more accurate and efficient than the set of all nodes, this result does not imply that any nodes in the lower state provide a good early warning signal. To show this, we investigated early warning signals constructed from half of the lower-state nodes whose $R_i$ score is the lowest---those with relatively few neighbors or few neighbors in the upper state. This node set, termed ``Lower Half'' and shown by the diamonds in Fig.~\ref{fig:examples-taus}, typically yielded lower $\tau$ values than the All, Lower State, and High Input node sets. This result implies that one needs to assemble an early warning signal from carefully chosen lower-state nodes such as those with large $R_i$ values. Finally, Upper State (shown by the crosses in Fig.~\ref{fig:examples-taus}) and Random (shown by the inverted triangles) node sets are either negatively correlated or not correlated with $D$, reinforcing our claim that the choice of nodes to be observed is essential. In sum, our simulation results suggest that, with a proper choice of observed node set---including the case of observing all nodes---standard multivariate and aggregated univariate indicators reliably increased in value prior to several transitions of nodes from the lower to upper state, performing well throughout a multistage transition.

\subsection{Robustness against Variation in Networks and Parameter Values}

\begin{figure}
    \centering
    \includegraphics[width = .7\textwidth]{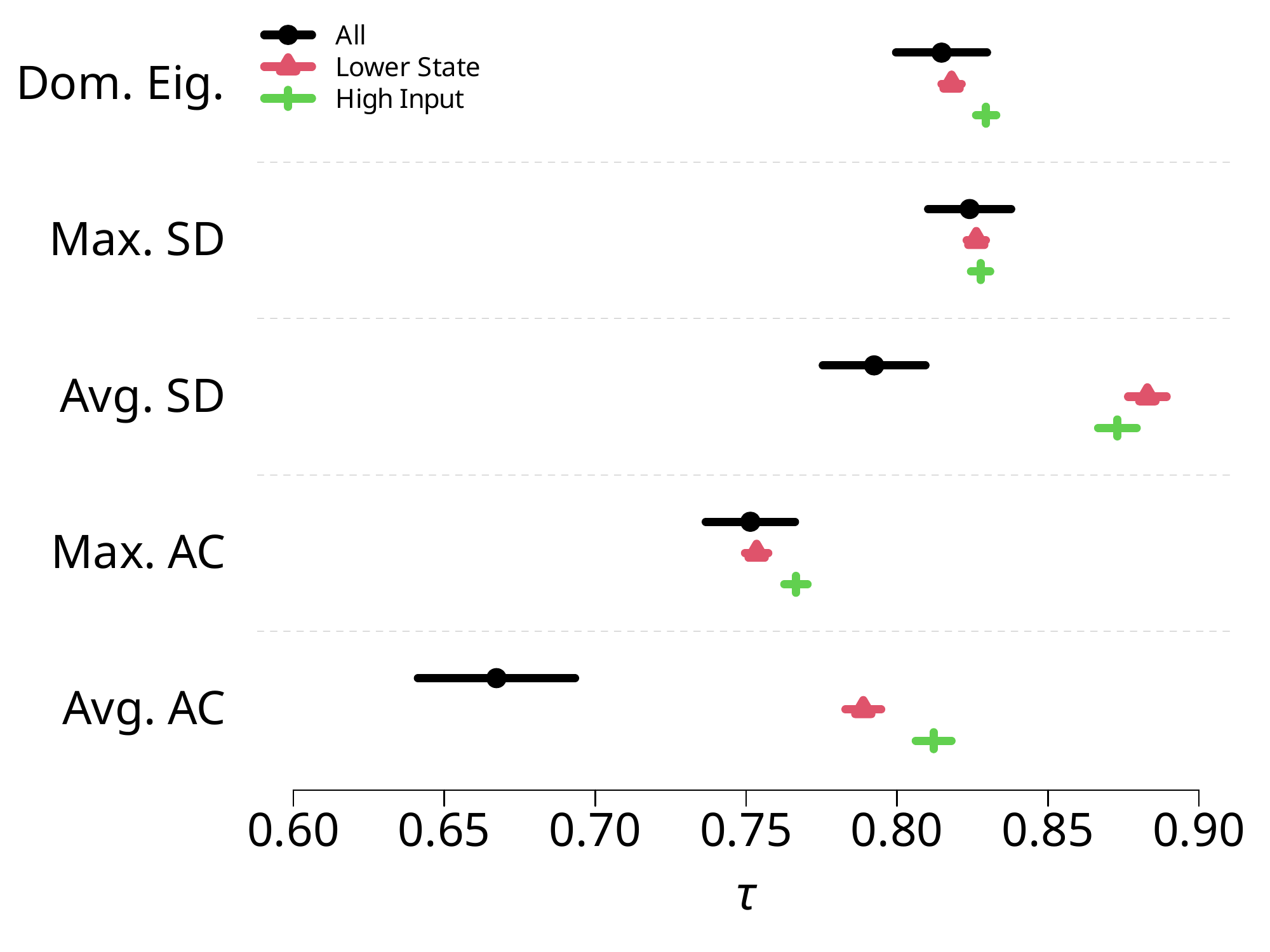}
    \caption{Predicted Kendall correlations ($\tau$) for five early warning signals and three node sets, estimated by a linear mixed effects model with a fixed effect for node set and a random effect for network. The results are based on the ten networks that have multiple stable ranges of $D$ in our numerical simulations. Markers (All: circles, Lower State: triangles, High Input: pluses) signify the predicted $\tau$ value. The horizontal lines represent the 95\% confidence intervals.}
    \label{fig:corrfig}
\end{figure}

To quantitatively examine the dependence of $\tau$ on network structure, we constructed a linear mixed effects model explaining $\tau$ with a fixed effect of node set and a random effect of network (Fig.~\ref{fig:corrfig}; see section \ref{sec:lme-table} for the statistical results).
We found that the predicted $\tau$ is large (i.e., approximately larger than $0.75$) across most networks, early warning signals, and node sets; the combination of the All node set and the average autocorrelation early warning signal yielded a somewhat lower predicted $\tau$ value (i.e., $0.667$). Variance-based methods (i.e., dominant eigenvalue and the maximum and average node-level standard deviation) tended to produce higher predicted $\tau$, ranging between $0.792$ and $0.828$. The autocorrelation methods produced lower predicted $\tau$, ranging between $0.667$ and $0.766$, although these values were still relatively high compared to other published results (e.g., \cite{dakos2008,dakos2012}).
The early warning signals based on the Lower State nodes were either no different (dominant eigenvalue, $p = 0.050$; maximum standard deviation, $p = 0.173$; and maximum autocorrelation, $p = 0.290$; uncorrected for multiple comparison) or better (average standard deviation, $p < 10^{-4}$; and average autocorrelation, $p < 10^{-4}$) than those based on all nodes. The early warning signals based on the High Input nodes improved over those based on all nodes ($p < 10^{-4}$ for all the early warning signals except the maximum standard deviation, for which $p = 0.025$) on average but were not as good as those based on the Lower State nodes in the case of the average standard deviation (High Input: $\tau = 0.873$, Lower State: $\tau = 0.883$). The $\tau$ values at most moderately depended on the network structure. Specifically, the distribution of random intercepts for network had the smallest standard deviation in the estimated linear mixed effects models for the maximum standard deviation early warning signal (0.022, 2.7\% of the magnitude of the intercept) and the largest standard deviation for the average autocorrelation early warning signal (0.041, 6.2\%).
These results are consistent with and generalize in terms of the variety of networks those shown in Fig.~\ref{fig:examples-taus}.

We then investigated the robustness of the results shown in Fig.~\ref{fig:examples-taus} against changes in parameter values. The full results are shown in the SI (see section \ref{sec:param-var}).
Consistent with previous results (e.g., \cite{clements2015}), decreasing the number of samples for calculating the early warning signal, $M$, has the strongest negative effect on the performance of early warning signals. We have also found that the average standard autocorrelation calculated from all nodes tends to perform worse than that calculated from the other node sets when the double-well equilibrium points are relatively close together (i.e., $(r_1, r_2, r_3) = (1, 3, 5)$ as opposed to $(1, 4, 7)$) or $r_1$, $r_2$, and $r_3$ are not evenly spaced (i.e.,  $(r_1, r_2, r_3) = (1, 2.5, 7)$ or $(1, 5.5, 7)$ as opposed to $(1, 4, 7)$).
As expected, allowing more than 50 TU for the model to relax to an equilibrium does not markedly improve the performance of the early warning signals. Thus, with the notable exception of the effect of $M$, the performance of each early warning signal is in general fairly similar across the different parameter settings.

\subsection{When We Do Not Know the Network Structure}

When we do not know the network structure, we cannot calculate $R_i$, which uses the adjacency matrix, to identify High Input nodes. Therefore, we explored the use of a correlation-based index, $R'_i$ (see section \ref{sub:node-set} for the definition), to choose alternative sentinel nodes, called the Large Correlation nodes, and computed the same set of early warning signals. We show the results for the Large Correlation node set by the box-times symbols in Fig.~\ref{fig:examples-taus}. The Large Correlation node set performed worse than the High Input node set. This result is expected because High Input uses the information about the network structure, whereas Large Correlation does not. However, the Large Correlation node set performed better than the Lower Half and Random node sets. In fact, $\tau$ with the Large Correlation node set is reasonably large in the dolphin network, roughly ranging between $0.6$ and $0.8$, whereas it is low in the power-law network (i.e., $\tau < 0.5$).
The discrepancy between the results for the two networks is associated with the different fidelity with which the Pearson correlation matrix, $\text{cor}(x_i, x_j)$, reflects the actual adjacency matrix (see Fig.~\ref{fig:threecorr}).

We also considered the nodes with the largest standard deviation in $x_i$, called Large SD, as another node set that does not need the information about the network structure. The rationale behind Large SD is that, when the $i$th node receives large input from other nodes, i.e., when $R_i$ is large, the standard deviation of $R_i$ should also be large because each $x_j$ in Eq.~\eqref{eq:doublewell} is fluctuating due to dynamical noise. A large fluctuation in $R_i$ is expected to make the standard deviation of $x_i$ large through Eq.~\eqref{eq:doublewell}. We found that early warning signals based on Large SD nodes (shown by stars in Fig.~\ref{fig:examples-taus}) perform better than those based on Large Correlation nodes and that the Large SD node set is approximately as well as the High Input node set. 
Both the Large SD and, to a lesser extent, the Large Correlation node sets perform well even when we consider all node transitions, not just those occurring after a stable range (Fig~\ref{fig:errors}). However, the Large SD node set is particularly sensitive to the number of samples used to determine node membership; its performance declines substantially on this test when we use only the first ten samples to determine node membership (Fig.~\ref{fig:sampling}).

Overall, these results support the idea of network-aware choice of sentinel nodes for early warning multistage transitions even when we do not have connectivity data at hand.

\subsection{Transition from the Upper State to the Lower State}
\label{sub:upper}

Simulations of Eq.~\eqref{eq:alt} on the power-law and dolphin networks with all nodes beginning in the upper state also show multistate transitions (see Fig~\ref{fig:examples-upper}). 
With Eq.~\eqref{eq:alt}, high-degree nodes receive a large positive contribution from the coupling term, which is the same as with Eq.~\eqref{eq:doublewell}. Therefore, lower-degree nodes or those adjacent to fewer upper-state nodes are most likely to transition from the upper to the lower state when $D$ gradually decreases. For this reason, Lower State and High Input, which are two node sets that performed well when we attempted to anticipate transition from the lower to upper states, are not expected to be equally good sentinels when the tipping direction is reversed, that is, when the system begins with nodes at the upper state and transits to the lower state.
Therefore, we additionally considered two node sets that are mirror images of Lower State and High Input. One is the set of nodes in the upper state, which we already considered in Fig.~\ref{fig:examples-taus}. The other is Low Input, which is the $n$ nodes with the smallest $R_i$ among the upper-state nodes; they are candidate of nodes that may transit from the upper to the lower state earlier than other nodes as $D$ decreases.

\begin{figure}
\centering
\includegraphics[width = \textwidth]{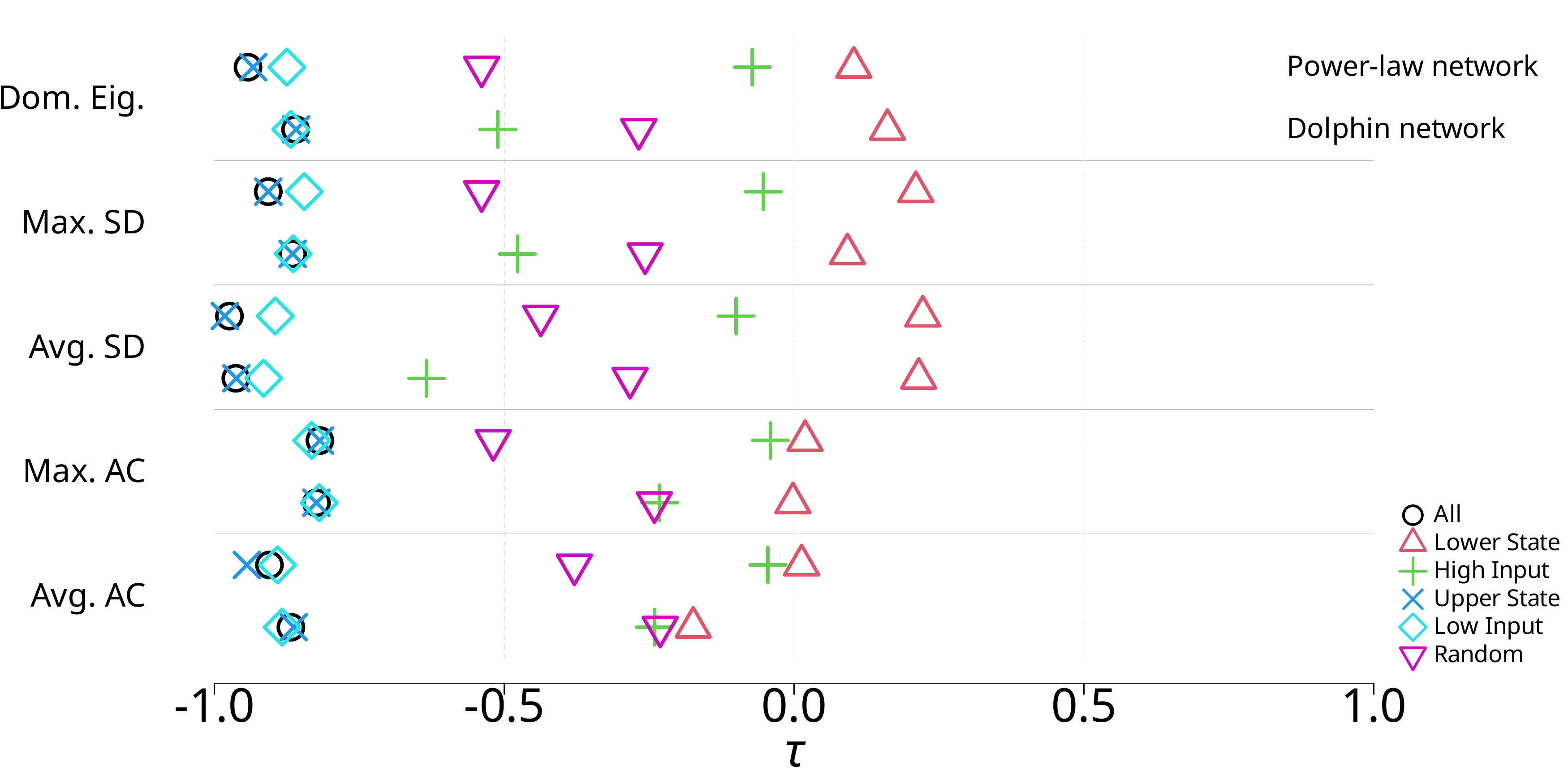}
\caption{Early warning signals in multistage transitions from the upper to lower equilibria. Kendall correlations ($\tau$) between each of the five early warning signals and the coupling strength, $D$, for different sets of nodes when the dynamics begin with the nodes in the upper state and $D$ gradually decreases are shown. 
See the caption of Fig.~\ref{fig:examples-taus} for the abbreviation of the early warning signals.
}
  \label{fig:examples-upper-taus}
\end{figure}

We show the Kendall's $\tau$ for the power-law and dolphin networks in Fig.~\ref{fig:examples-upper-taus}.
In Fig.~\ref{fig:examples-upper-taus}, a negative $\tau$ indicates that the early warning signal became large as $D$ decreased towards a transition from the upper to the lower state. Therefore, large negative $\tau$ values are indicative of critical slowing down as we decrease $D$. We find that the early warning signals calculated from lower-state nodes (Lower State, shown by the triangles, and High Input, shown by pluses) are not useful for anticipating transitions. In contrast, those calculated from the All node set (shown by the circles) or those informed by upper-state node dynamics (Upper State, shown by crosses; Low Input, shown by diamonds) are highly negatively correlated with $D$. This result indicates that the nodes in the upper state, not those in the lower state, provide useful early warning signals. Furthermore, the best sentinel nodes are opposite in terms of $R_i$ from when we started with the lower equilibrium and observed transitions of the nodes from the lower to the upper state. A suitable choice of sentinel nodes depends on the tipping direction, even if the dynamical system model is similar or essentially the same.

\section{Discussion} 
\label{sec:discussion}

We showed that both multivariate (i.e., eigenvalue-based) and aggregated univariate (i.e., variance- and autocorrelation-based)
early warning signals can provide advance notice of state changes in multistage transitions in coupled double-well systems. 
Furthermore, we showed that constructing early warning signals only based on a subset of nodes, called sentinel nodes, is competitive with, and sometimes more effective than, using all nodes to calculate the early warning signals. Specifically, it is useful to monitor nodes that have not transitioned to the alternative state but are connected to other nodes that have already transitioned to such a state. We showed that the early warning signals calculated based on the thus selected sentinel nodes were effective both when nodes were transitioning from a lower state to an upper state and vice versa. Up to our numerical efforts, the results were robust against parameter variation, network structure, and choice of early warning signals.

We have shown that the choice of which nodes to monitor for early warning signals has a marked impact on the effectiveness of the early warning signal. In particular, when we observed transitions from the lower to upper states, a good set of nodes to monitor was those with a large degree or with many connections to other nodes that have already transitioned to the upper state, as quantified by $R_i$.
At first glance, this result seems at odds with those by Aparicio et al. \cite{aparicio2021}, who used network control theory to propose that lower-degree nodes tended to make better sentinels. In fact, in their model, the dynamics always starts with nodes in the upper state because it is a model of species abundance and its loss.
We showed that lower-degree nodes are good sentinel nodes when the nodes are initially in their upper states and transit to their lower states as a bifurcation parameter gradually changes. Aparicio et al. provided two indices for the suitability of their sentinel nodes. Because one of the two indices only depends on the network structure, we calculated the other measure, called $\rho$, for our simulations given the network. A value of $\rho$ closer to zero indicates that their sentinel nodes are more suitable. We found for our power-law network
$\rho = 0.042$ when all nodes start in the lower state and $\rho = 0.038$ when all nodes start in the upper state;
for the dolphin network, we obtained $\rho = 0.030$ and $\rho = 0.007$, respectively. These results are consistent with our numerical results, in which low-degree nodes provide informative early warning signals when we started with the upper but not the lower state. We emphasize that a good choice of sentinel nodes depends on the initial condition and the tipping direction even if we fix the dynamical system as well as the network structure.

There are many cases in which a network model is thought to represent a complex system showing tipping phenomena but the edges of the network are not directly known \cite{brugere2018}. Examples include the co-occurrence of symptoms of neurological conditions \cite{hofmann2016} and the rates of return on traded financial securities \cite{quax2013}. In such cases, we are typically given only multivariate time series data and want to derive informative early warning signals for tipping points that possibly constitute a multistage transition.
A strategy in this situation is to infer the network structure from multivariate time series data \cite{brugere2018,timme2014} and then calculate candidate sentinel nodes from the estimated network using, for example, the node's ranking in terms of $R_i$.
We avoided this approach because network inference from time series data is subject to error due to, e.g., thresholding decisions \cite{brugere2018} or uncertainty in model estimation \cite{timme2014}. Instead, we proposed a method to identify sentinel nodes only based on the Pearson correlation between 
the time series at pairs of nodes, which provides a proxy to edges (although one should not use the Pearson correlation as an estimate of the network edge in general \cite{zalesky2012}).
Our sentinel nodes determined based on the Pearson correlation provided reasonably strong early warning signals, but their performance did not reach that for the case in which we know the network structure. However, choosing sentinel nodes based on the standard deviation of the node's state performed in a similar manner to sentinel nodes chosen using information on network structure.
Finding better sentinel nodes given multivariate time series data for which the explicit network structure is unknown warrants future work.
We also point out that we currently do not have equivalent methods when the nodes are initially in their upper states and transit to their lower states as the value of a control parameter gradually varies, which is typical in ecological modeling.

Although we have shown that High and Low Input node sets are efficient at anticipating major changes of state in the models we studied, there is much room for further improvements. First, multistage transitions imply that there are intermediate stages in which some nodes have tipped and the others have not and that we have seen a history of which nodes have tipped and when. If we use such information, we may be able to improve performances of early warning signals with respect to both the node set selection and the definition of the signal. Second, it may be helpful to use benchmark networks that show multistage transitions.
If a network is composed of multiple disconnected components of tipping elements, the entire network should show multistage transitions because the different disconnected components show a bifurcation at different values of a control parameter in general. Therefore, a network with a strong planted community structure is expected to show multistage transitions for various dynamical systems.
Degree-heterogeneous random graphs also show multistage transitions, which is underpinned by both numerical simulations and a mean field theory
\cite{kundu2022b}. Studying multistage transitions and early warning signals on these networks may be useful.

We used cubic polynomials to drive the node's dynamics (and hence a potential in the form of quartic polynomials) and unipartite networks to test our ideas. These modeling assumptions are reasonable for 
investigating, for example, climate and vegetation cover transitions \cite{wunderling2021,wunderling2022}. In contrast, various ecological systems are better modeled by bipartite networks, in which
the two layers of nodes typically represent pollinators (or seed dispersers) and plants \cite{lever2020,bascompte2007}. In fact, ecological dynamics on bipartite networks also show multistage transitions \cite{lever2020}.
Despite the seminal work based on network control theory \cite{aparicio2021}, discussed above, further work is desirable for identifying informative sentinel nodes in ecological dynamics on bipartite networks.
Other types of dynamics such as reactive and synchronization dynamics on networks should also be investigated.
Additionally, although saddle-node bifurcations have been frequently studied, natural systems may also show other types of bifurcations. Early warning signals for transcritical, Hopf, and other bifurcations are beyond the scope of this work, but anticipating such transitions is important in several fields, including the epidemiology \cite{drake2019} and ecology \cite{lever2020}.
Finally, although we have shown that a careful choice of sentinel nodes can dramatically reduce the amount of data needed without sacrificing the quality of early warning signals, we are ignorant of the amount of the data needed from each node in this study. Shortening the length of temporal data required will be an important next step, given that sampling can be expensive and invasive in various applications such as ecology and medicine. Spatial correlations such as Moran's $I$ have been used to provide early warning signals on square lattices \cite{dakos2010}, and their extensions to the case of complex networks may help reduce the required amount of temporal sampling.

In addition to sampling limitations, the specificity of early warning signals is a known challenge \cite{gsell2016,boettiger2012,bury2021,boerlijst2013}.
Suppose that an early warning signal tends to increase as a control parameter gradually increases towards a tipping point.
It is difficult in general, however, to suggest a particular range of values of the early warning signal that indicates an impending transition.
In fact, the Kendall's $\tau$, which is deemed to be a standard performance measure, may be large for several reasons, including when the early warning signal monotonically increases as the control parameter increases regardless of tipping points \cite{boettiger2012}.
This lack of specificity is also present in our results (see Fig. ~\ref{fig:examples}).
Developing methods, such as maximum likelihood \cite{boettiger2012} or algorithmic classification \cite{bury2021} techniques, to improve the specificity of early warning signals is an important area of further research.
With all these tasks saved for future work, by combining information about the network structure and dynamics, the present study takes a significant step towards accurately and cost-efficiently anticipating different types of tipping points in complex dynamical systems.

\section*{Data accessibility}

The datasets generated and analyzed during the current study are available in the GitHub repository, \href{https://github.com/ngmaclaren/doublewells}{https://github.com/ngmaclaren/doublewells}, along with all relevant computer code.

\section*{Acknowledgments}

We thank Hiroshi Kori and Makito Oku for valuable discussion.

\section*{Author contributions}

N.M. conceived and supervised the project. N.G.M. performed the simulations and computations with assistance from P.K.. N.G.M. and N.M. analyzed the data and wrote the paper.

\section*{Funding}

N. Masuda acknowledges support from AFOSR European Office (under Grant No. FA9550-19-1-7024), the Sumitomo Foundation, the Japan Science and Technology Agency (JST) Moonshot R\&D (under Grant No. JPMJMS2021), and the National Science Foundation (under Grant No. 2052720).

\section*{Conflict of interest declaration}

The authors declare no competing interests.

\bibliographystyle{unsrt}
\bibliography{refs.bib}


\appendix

\section*{Supplementary Information}

\subsection{Choice of the threshold for classifying the nodes into their lower and upper states}
\label{sec:demo}

In the absence of a coupling term, the double well model $\frac{dx}{dt} = -(x - r_1)(x - r_2)(x - r_3)$ with $(r_1, r_2, r_3) = (1, 4, 7)$ has inflection points at $x_i \approx 2.268$ and $x_i \approx 5.732$. Values of $x$ near $r_2$---that is, $2.268 < x < 5.732$---are transient and are not expected in a system at equilibrium. In this section, we demonstrate that these inflection points provide sensible thresholds for classifying nodes as either in a lower or upper state even in the presence of a coupling term. 

\begin{figure}
  \includegraphics[width = \textwidth]{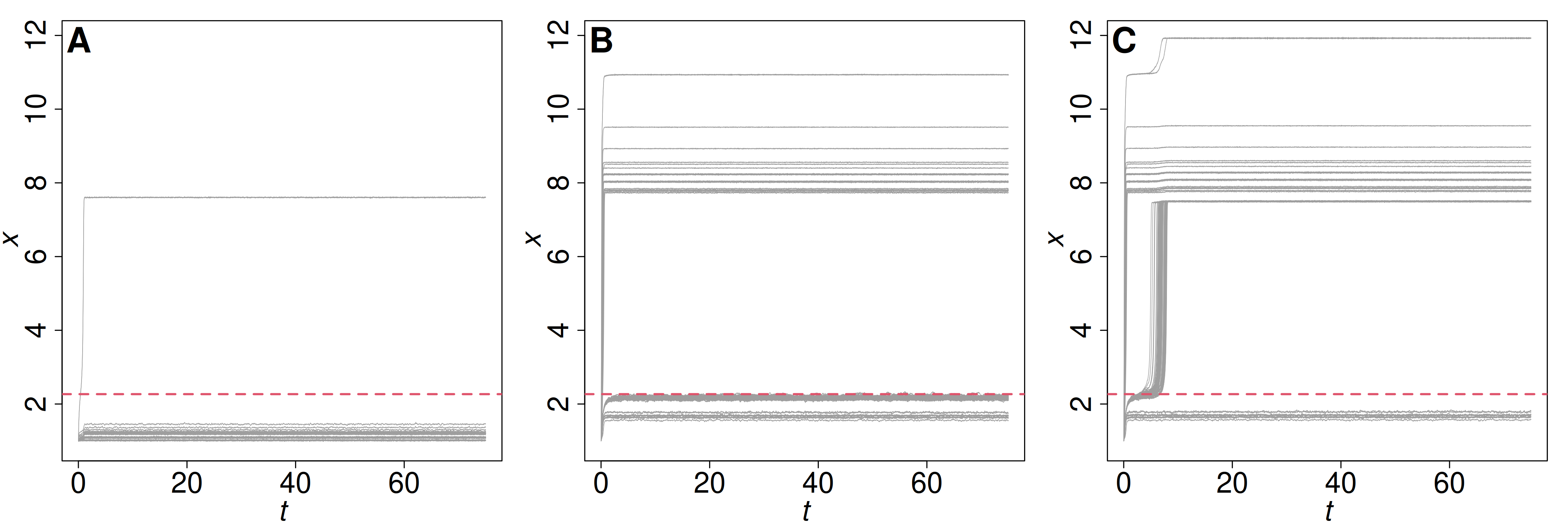}
  \caption{Example simulations for $D = 0.20$ (A), $D = 0.945$ (B), and $D = 0.950$ (C). The gray lines show the state $x_i$ of each node. The dashed line marks the lower inflection point for the double-well model in the absence of coupling.}
  \label{fig:demo}
\end{figure}

Figure~\ref{fig:demo} shows time courses of $x_i$ $\forall i \in \{1, \ldots, N\}$ as a function of time for three values of $D$. We carried out the simulations on the power-law network used as an example in the main text (see Fig.~\ref{fig:examples}A) using our standard parameter values and initializing the nodes near the lower stable point (see Section~\ref{sub:sims} for the details). 

In Fig.~\ref{fig:demo}A, the system is far from a transition (we set $D = 0.200$; the next transition is at $D = 0.300$), and most nodes (i.e., 98 nodes) remain near the lower stable point. The threshold used in the main text, i.e., $x = 0.268$, correctly and clearly partitions all the nodes into their lower and upper states although the $x_i$ values at the lower or upper state depend on the node. When we placed the bifurcation parameter near a transition (i.e., $D=0.945$; see Fig.~\ref{fig:demo}B), as in Fig.~\ref{fig:demo}A, the state variables, $x_i$, rapidly converge to their equilibrium positions and remain near there. Now, 29 nodes are apparently in the upper state. Of the other 71 nodes, some nodes are close to the threshold but their $x_i$ values are lower than the threshold except that some of them are briefly above the threshold due to dynamical noise. At $D=0.950$, most nodes (i.e., 89) are in the upper state although it now takes more time to converge (see Fig.~\ref{fig:demo}C). With this $D$ value, the threshold clearly distinguishes between the lower and upper states for all nodes, which is similar to Fig.~\ref{fig:demo}A. In sum, we conclude that the threshold we used, or in practice threshold values satisfying $2.268 < x < 5.732$, is reliable in classifying the nodes in the coupled double-well dynamics into the lower and upper states.

\subsection{Networks}
\label{sec:networks}

We used the following five model networks and 17 empirical networks. We coerced all networks to be simple, undirected, and unweighted; if multiple components were present, we only used the largest connected component. Of these networks, 10 networks had clear multistage transitions: the Dolphin, Elephant Seal, Karate, Nestbox, PIRA, Tortoise, Vole, and Weaverbird empirical networks and the Fitness and Powerlaw model networks.

\subsubsection{Model Networks}
\label{sub:modelnetworks}

Model networks were generated to have an expected $N = 100$ nodes and an edge density of approximately 0.05. These values were arbitrary but generally reflect the observation that many empirical networks for which early warning signals and resilience have been discussed are relatively small and sparse. For each model network, we chose parameter values that produced networks that were reasonably close to these expectations.

\begin{description}
\item [Erd\H{o}s-R\'{e}nyi random graph] We constructed a network using the Erd\H{o}s-R\'{e}nyi (ER) random graph \cite{erdos1959} with $N = 100$ nodes. We set the probability of an edge between each pair of nodes to $0.05$. The generated network was a connected network with 246 edges.

\item [Random graph with communities] This variation on the ER random graph has a planted community structure \cite{csardi2006}. We generated a network with $N = 100$ nodes grouped into five networks of equal size (i.e., $20$ nodes), each of which we generated as an ER random graph. The nodes in each subgraph were adjacent to each other independently with probability $p = 0.2475$. Each pair of subgraphs was adjacent by exactly one edge, whose endpoints we selected uniformly at random from each of the two subgraphs. With this $p$ value, the expected edge density of the generated network is equal to that of the aforementioned ER random graph. The generated network had one disconnected node; after removing that node the network had 99 nodes and 244 edges.

\item [Barab\'asi-Albert model] We used a Barab\'asi-Albert (BA) network  \cite{barabasi1999} with $N = 100$ nodes. We set the number of edges that each new node has, denoted by $m$, to $2$. We assumed that the initial network in the network growth process was a graph with two nodes and one edge connecting them. As a result, the generated network had 197 edges. The BA network produces the power-law degree distribution with power-law exponent 3 in the limit of $N\to\infty$.

\item [LFR model] The Lancichinetti-Fortunato-Radicchi (LFR) benchmark model generates networks with an explicit community structure and both node degree and community size being drawn from power-law distributions \cite{lancichinetti2008}. We used a network with $N = 100$ nodes. We set the power-law exponent of the expected degree distribution to $-2$ and that of the community size distribution to $-1.5$. Additionally, we set the average degree to $5$, the maximum degree to 20, the maximum community size to 20, the minimum community size to 10, and the probability that an edge connected nodes in different communities to $0.1$. The generated network had 269 edges. The generated network was a connected network. We used the implementation of this model in the NetworkX (v2.7.1) package \cite{hagberg2008} for Python (v3.10.2).

\item [Fitness model] As an alternative to the BA model, we generated a network with a heterogeneous degree distribution using a node fitness model \cite{goh2001}. We used an expected power-law degree distribution with power-law exponent $-2$. To accomplish this, we set the fitness of node $i$ to $(i + i_0 -1)^{-\alpha}$, where $\alpha = 1$ and $i_0$ is a constant that constrains the maximum node degree \cite{chung2002}. Following Cho et al \cite{cho2009}, we set $i_0 = N^{1-\frac{1}{\alpha}}(10\sqrt{2}(1 - \alpha))^{\frac{1}{\alpha}}$. As a result, the degree distribution is theoretically a power-law distribution with power-law exponent $-2$ \cite{goh2001}. We generated a network with 100 nodes and 197 edges. Thirteen nodes were isolated, and the other nodes formed a single connected component. Therefore,
the largest connected component contained $N=87$ nodes and 197 edges.

\item [Configuration model with a power-law degree distribution] We generated a network with a power-law degree distribution using a configuration model. We set $N=100$. To this end, we sampled the degree $k$ of each node independently from a discrete version of the Pareto distribution given by $p(k) = \frac{k^{-\alpha}}{\zeta(\alpha, k_{\rm min})}$, with $k \in \{1, 2, \ldots\}$ and $\zeta(\alpha, k_{\rm min}) = \sum_{k'=k_{\min}}^{\infty} (k')^{-\alpha}$ \cite{clauset2009}. We set $\alpha = -2$ and $k_{\rm min} = 1$. Different configuration model algorithms can lead to different network structures \cite{fosdick2018}. We used the ``vl'' algorithm, attributed to Fabien Viger and Matthieu Latapy by \cite{csardi2006}, which attempts to sample uniformly from all the possible simple undirected graphs specified by the input degree distribution, under the constraint that the generated network is connected. The generated network had 146 edges.

\end{description}

\subsubsection{Empirical Networks}
\label{sub:empiricalnetworks}

All empirical networks were selected from the ``networkdata'' R package \cite{schoch2022}; a short name is given along with the name of the data object that stores the network in the ``networkdata'' package. Many networks published with the ``networkdata'' package are presented as lists of several networks of the same types. In this case, we chose the network from such a list most closely matching the number of nodes and edges in our model networks.

\begin{description}
\item [Bat] The first network in ``animal\_18'' \cite{silvis2014}. Two bats ({\em Myotis sodalis}) are adjacent if they shared the same roost on the same day. This network is a single-mode projection of the original bipartite network. This network has 43 nodes and 546 edges.
\item [Dolphin] The ``dolphins\_2'' network \cite{lusseau2003}. Two dolphins are adjacent if they were observed together more frequently than expected by chance. There are 62 nodes and 159 edges in this network. 
\item [Drug user] The ``covert\_17'' network \cite{weeks2002}. Two individuals are adjacent if at least one individual named the other in a ``name generator'' sociometric survey; there are 193 nodes and 273 edges. 
\item [Elephant seal] The first network in ``animal\_23'' \cite{casey2015}. Two northern elephant seals ({\em Mirounga angustirostris}) are adjacent based on observed dominance behaviors during a given day. This network has 46 nodes and 56 edges.
\item [Hall] The ``hall'' network \cite{freeman1998}. Two individuals residing in the surveyed university residence hall are adjacent if at least one nominated the other as a friend in a sociometric survey. This network has 217 nodes 1,839 edges.
\item [Highschool boy] The ``highschool\_boys'' network \cite{coleman1964}. Two individuals are adjacent if at least one nominated the other as a friend in either of the two sociometric survey waves. This network has 70 nodes and 274 edges.
\item [House finch] The ``animal\_6'' network \cite{adelman2015}. Two house finches ({\em Haemorhous mexicanus}) were connected if they used the same bird feeder in a given day. This network has 108 nodes and 1,026 edges.
\item [Jazz] The ``jazz'' network \cite{gleiser2003}. The nodes in this network are jazz bands. Two nodes are connected if at least one musician played in both bands. This network has 198 nodes and 2,742 edges.
\item [Karate] The so-called karate club network, ``karate'' \cite{zachary1977}. Two individuals are adjacent if they had a social relationship outside of the club. This network has 34 nodes and 78 edges.
\item [Lizard] The first network in ``animal\_36'' \cite{bull2012}. Two sleepy lizards ({\em Tiliqua rugosa}) are adjacent if GPS tags recorded them within two meters of each other during a ten-minute interval on a given day. This network has 60 nodes and 318 edges.
\item [Nestbox] The third network in ``animal\_11'' \cite{firth2015}. The nodes are individual birds of several species (\textit{i.e.}, great tits, {\em Parus major}, blue tits, {\em Cyanistes caeruleus}, marsh tits, {\em Poecile palustris}, coal tits, {\em Periparus ater}, and Eurasian nuthatches, {\em Sitta europaea}). Two individuals are adjacent if they visited the same nest box on the same day. This network has 126 nodes and 1,615 edges.
\item [Netsci] The ``netsci'' network \cite{newman2006}. Two authors publishing in the field of network science are adjacent if they had published together at the time the data was gathered. This network has 379 nodes and 914 edges.
\item [PIRA] The first network in ``covert\_38'' \cite{gill2014}. Nodes in this network are individuals with a connection to the Provisional Irish Republican Army (PIRA). Two nodes are adjacent if they were both members of the PIRA in a given time period, or connected by a friendship or family relationship to a PIRA member. This network has 128 nodes and 201 edges.
\item [Surfer] The ``surfersb'' network \cite{freeman1988}. Two members of a windsurfing community are adjacent if they were observed together over a period of 31 days. This network has 43 nodes and 336 edges.
\item [Tortoise] The second network in ``animal\_35'' \cite{sah2016}. Two desert tortoises ({\em Gopherus agassizii}) are adjacent if they were observed to asynchronously use the same burrow within a given observation period. This network has 94 nodes and 181 edges. 
\item [Vole] The 88th network in ``animal\_33'' \cite{davis2015}. Two voles are adjacent if they were caught in at least one common trap over the observation period. This network has 111 nodes and 240 edges.
\item [Weaver bird] The 17th network in ``animal\_10'' \cite{vandijk2014}. Two weaver birds ({\em Philetairus socius}) were adjacent if they were observed to enter the same nest for any purpose over the observation period. This network has 64 nodes and 177 edges.
\end{description}

\subsection{Anticipating multistage transitions as we vary $u$}
\label{sec:stress}

We used the connection strength, $D$, as a bifurcation parameter in the main text. In this section, we consider stress applied to all nodes uniformly (i.e., $u$ in Eq.~2 in the main text), as the bifurcation parameter instead.
We set $D=0.1$. To observe the tipping from the lower to the upper state, we began each series of simulations with $u = 0$. We increased $u$ by 0.05 (i.e., $u~\in~\{0, 0.05, 0.10, \ldots\}$) for each simulation until $>90$\% of nodes were in the upper state at equilibrium.
At each value of $u$, we ran a simulation with all nodes starting in the lower state (i.e., $x_i = 1~\forall~i$) and measured the early warning signals. 
To observe the tipping from the upper to the lower state, we began each series of simulations with
$u = 0$ and decreased $u$ by 0.05 (i.e., $u~\in~\{0, -0.05, -0.10, \ldots\}$) for each simulation until $>90$\% of nodes were in the lower state at equilibrium.
At each value of $u$, we started a simulation with $x_i = 7~\forall~i$.

\begin{figure}
  \includegraphics[width = .6\textwidth]{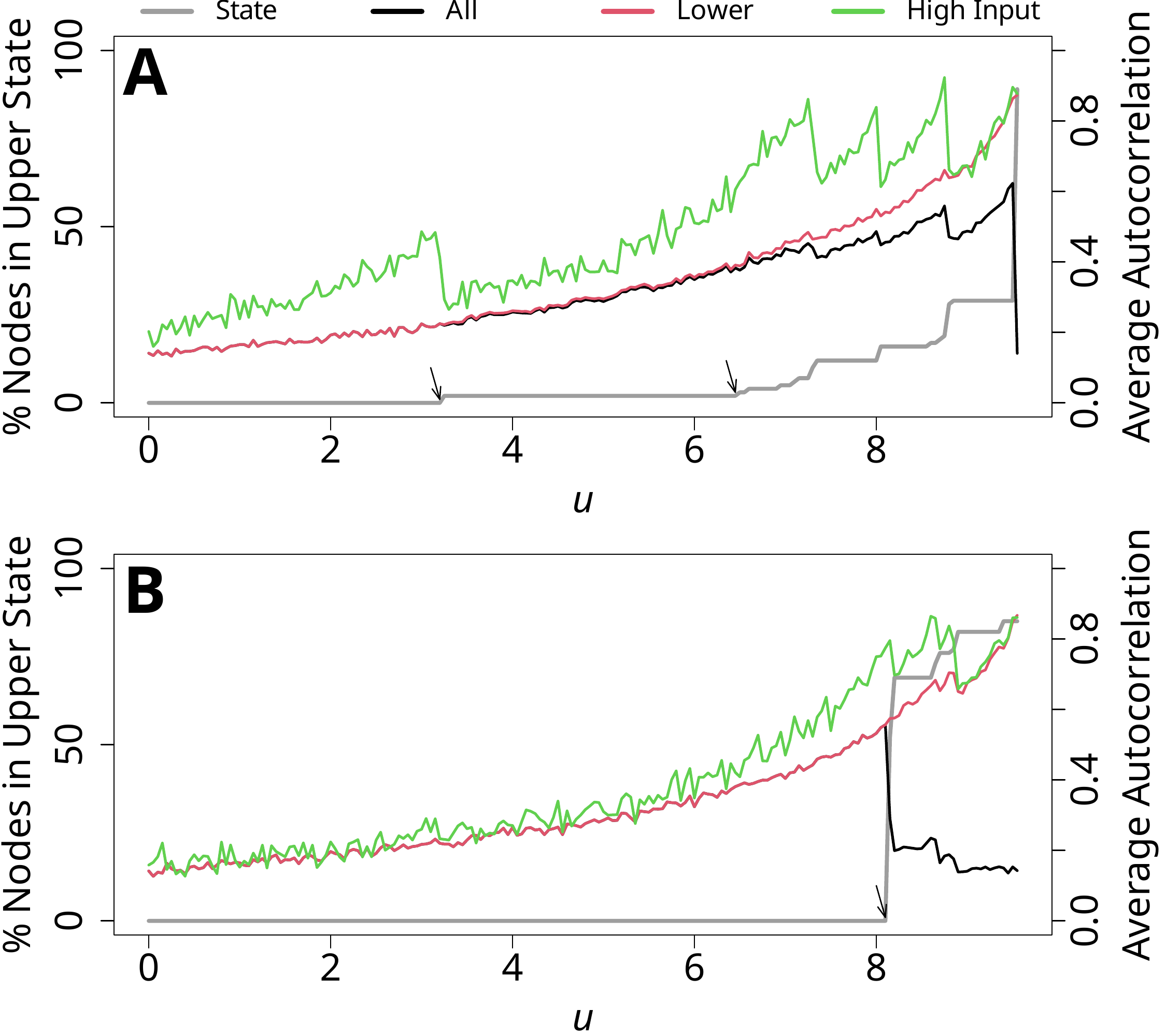}
  \caption{Multistage transitions when nodes are initially in the lower state and we vary $u$. (A) Power-law network. (B) Dolphin network.}
  \label{fig:u1}
\end{figure}

\begin{figure}
  \includegraphics[width = \textwidth]{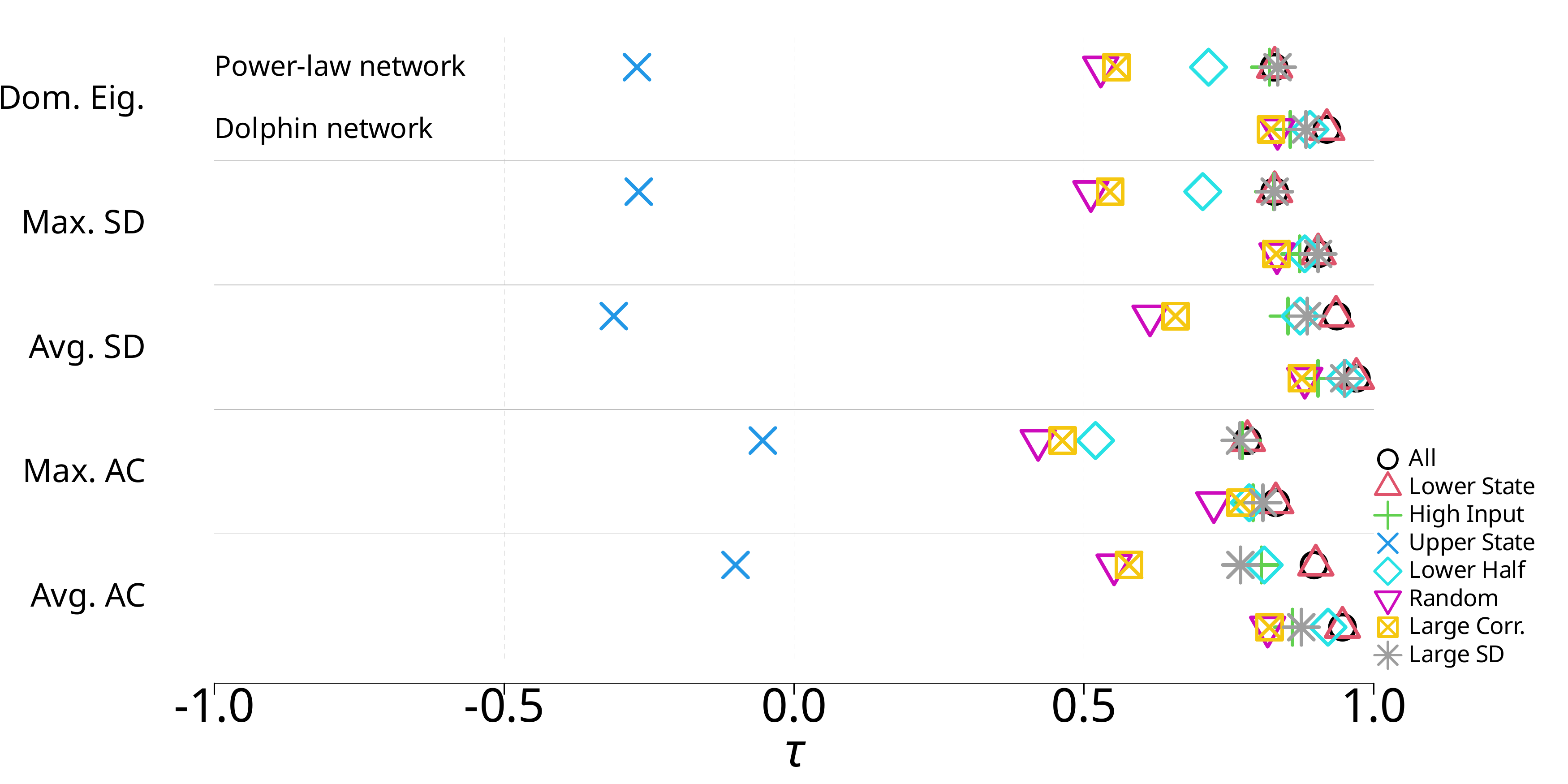}
  \caption{Kendall's $\tau$ between each of the five early warning signals and $u$ for the different node sets when nodes begin in the lower state and we gradually increase $u$. Dom. Eig.: dominant eigenvalue of the covariance matrix of all nodes in the node set; Max. SD, Avg. SD: maximum and average standard deviation of $x_i$; Max. AC and Avg AC: maximum and average autocorrelation of $x_i$; Large Corr. abbreviates the Large Correlation node set.}
  \label{fig:u2}
\end{figure}

Figure~\ref{fig:u1} shows that, for the two example networks, i.e., the power-law and dolphin networks, there are fewer stable ranges of $u$ than of $D$. However, a multistage transition is still evident for both networks when we vary $u$, whereas the dolphin network has much narrower stable ranges of $u$ after the first transition. Nonetheless, the early warning signals calculated on the basis of each of our primary node sets (i.e., All, Lower State, and High Input) increase in value prior to approaching bifurcation points (see Fig.~\ref{fig:u2}). An important difference between $D$ and $u$ as bifurcation parameter is that the High Input node set, while still displaying a large $\tau$, is no longer as good as or better than the All or Lower State nodes. However, the $\tau$ value for the High Input node set is still high ($> 0.7$).

\begin{figure}
  \includegraphics[width = .6\textwidth]{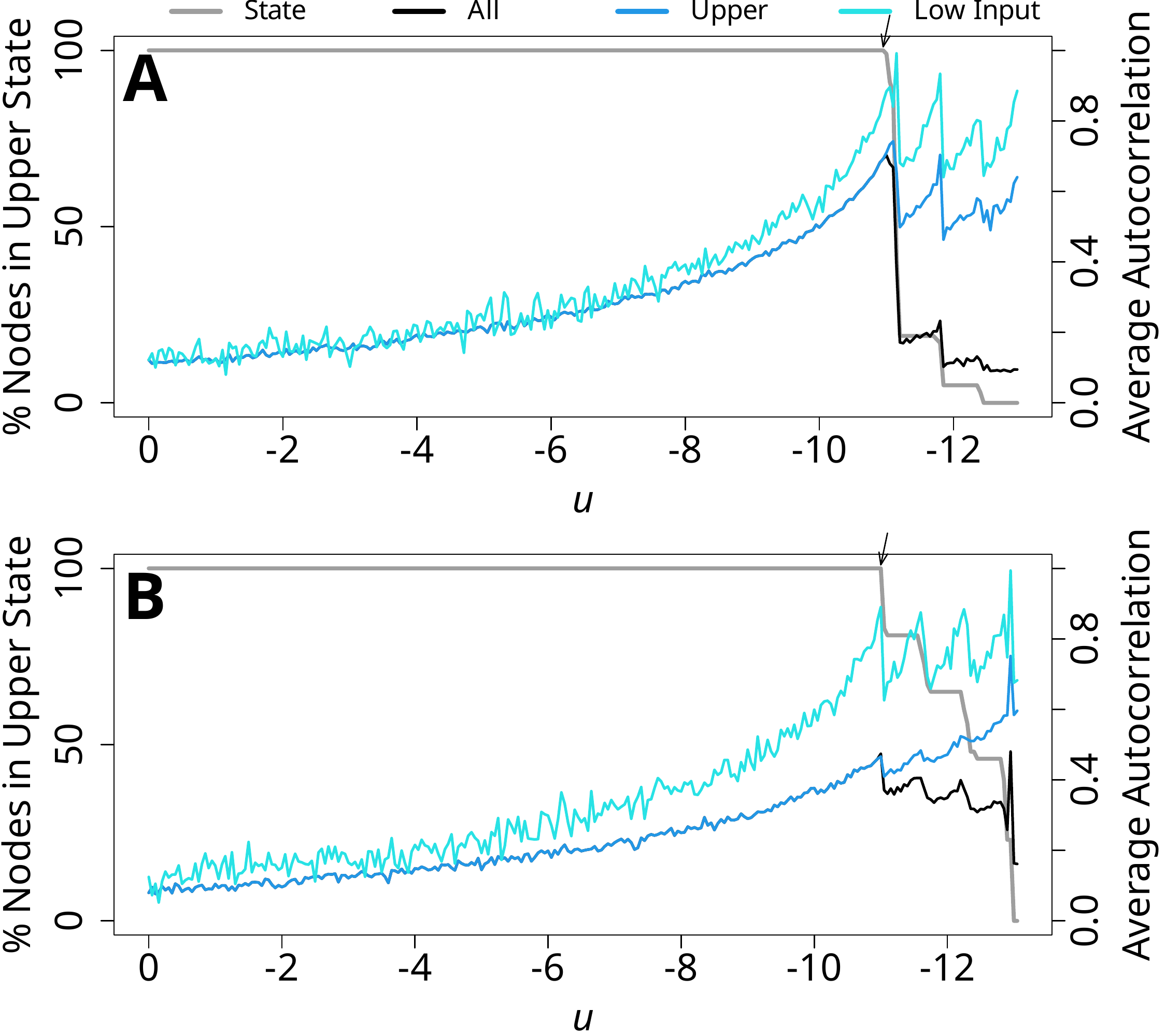}
  \caption{Multistage transitions when nodes are initially in the upper state and we gradually decrease $u$. (A) Power-law network. (B) Dolphin network.}
  \label{fig:u3}
\end{figure}

\begin{figure}
  \includegraphics[width = \textwidth]{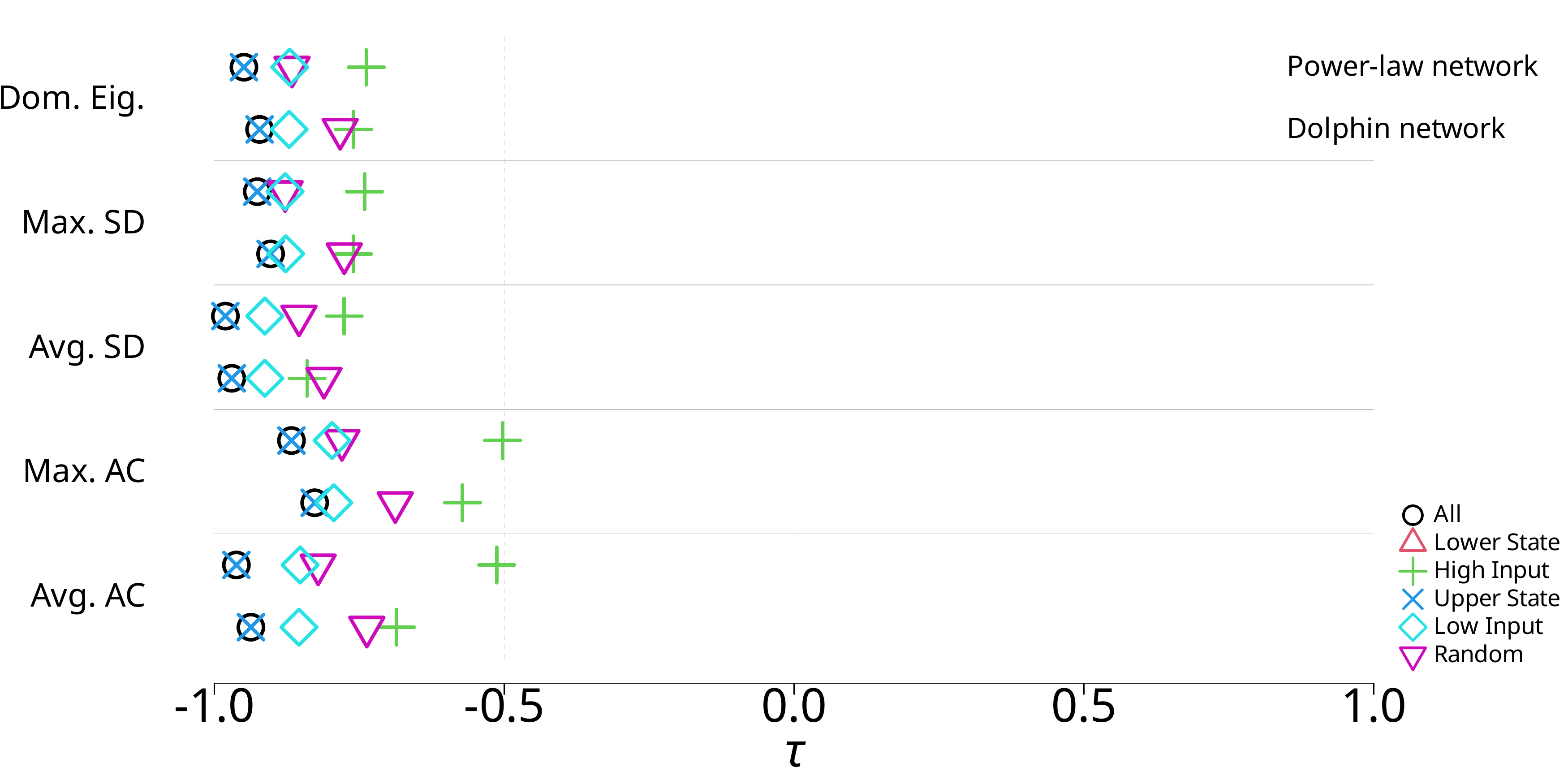}
  \caption{Kendall's $\tau$ between $u$ and each of the five early warning signals for different sets of nodes when nodes begin in the upper state and we gradually decrease $u$. See Fig.~\ref{fig:u2} for abbreviations.}
  \label{fig:u4}
\end{figure}

The differences between the two bifurcation parameters are stronger for the transition from upper state to lower state as we decrease $u$ (Figs.~\ref{fig:u3} and \ref{fig:u4}). We observe one primary stable range and several smaller stable ranges thereafter for both networks (Fig.~\ref{fig:u3}). Additionally, even the Random node set performs reasonably well in anticipating the transition. As expected, the High Input node set still performs notably worse than the Low Input node set.

\subsection{Analysis Considering Each Major Transition}
\label{sec:eachtransition}

Previous authors have noted that their early warning signal methods performed well during some part of a multistage transition but less well during others. For example, as we mentioned in the main text, Lever et al. \cite{lever2020} found that their method performed well early in a multistage transition, whereas Aparicio et al. \cite{aparicio2021} found that their method performed well late in a multistage transition. To investigate the performance of our method across different state-transition events constituting a multistage transition, we calculated the $\tau$ values for each of the major transitions, i.e., the transitions of some nodes occurring after a stable range of at least 15 unique values of $D$, in the power-law and dolphin networks. We separately analyzed the $\tau$ value for each major transition without averaging it over all the major transitions as was done in the main text.

\begin{figure}
  \includegraphics[width = \textwidth]{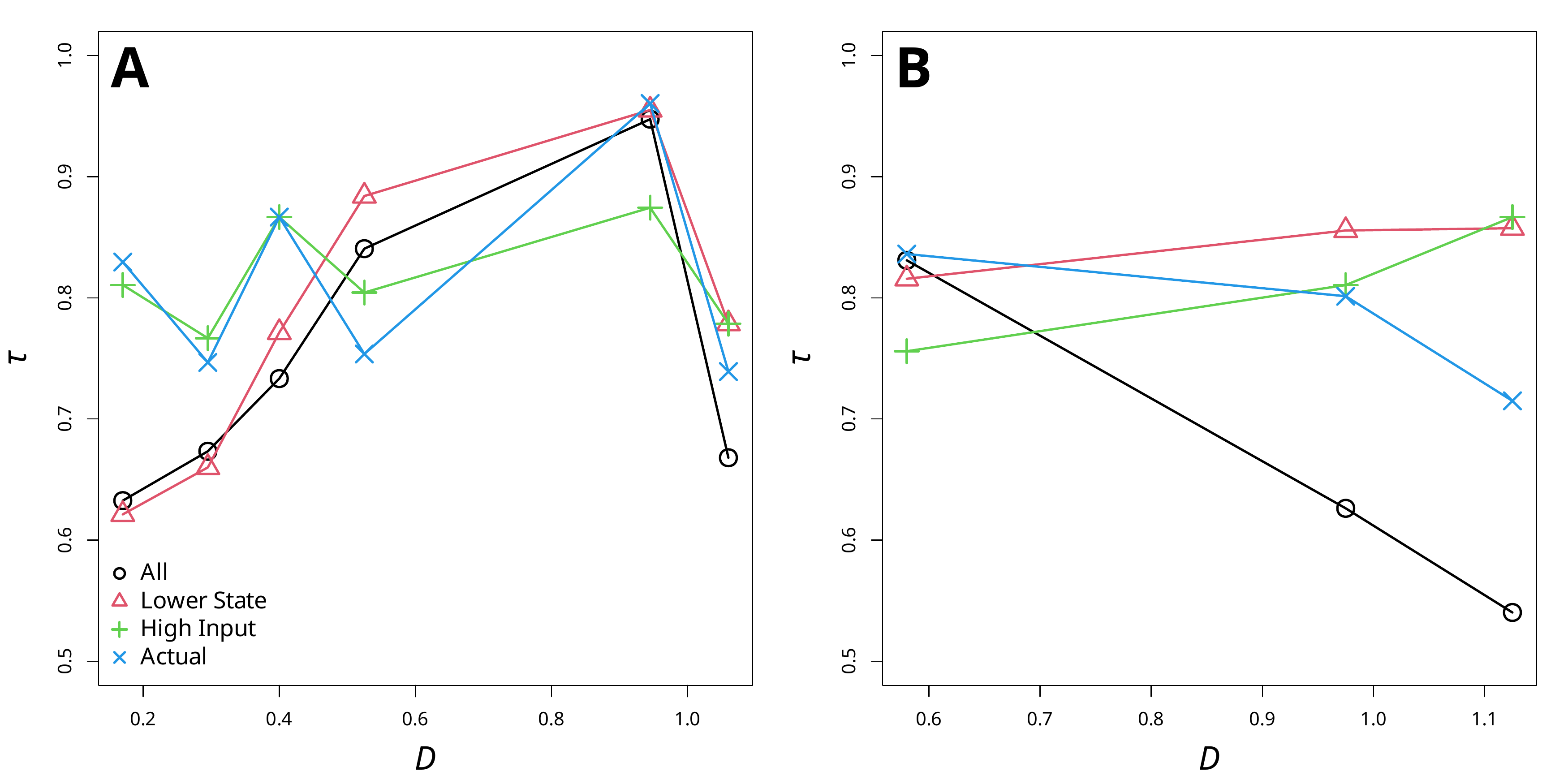}
  \caption{Kendall's $\tau$ for the average autocorrelation signal associated with each transition of some nodes after a stable range of $D$. Each symbol corresponds to a transition. We examined the four node sets, ``All'', ``Lower State'', ``High Input'', and ``Actual''. ``Actual'' is the set of the nodes that actually transitioned at the end of the stable range. (A) Power-law network. (B) Dolphin network.
  }
  \label{fig:each}
\end{figure}

In Fig.~\ref{fig:each}, we show Kendall's $\tau$ as a function of $D$ for each major transition in the power-law (Fig.~\ref{fig:each}A) and dolphin (Fig.~\ref{fig:each}B) networks. Each symbol corresponds to a major transition. For the power-law network, the All (circles) and Lower State (triangles) node sets perform less well in the first major transitions, better during the fourth and fifth major transitions, and less well for the last transition. In contrast, the High Input node set (pluses) performs roughly equally well in the different transitions. For the dolphin network, there are only three transitions. The High Input node set again performs somewhat consistently across these three transitions. The All node set declines in performance across the three transitions. For both networks, the High Input node set performs well, with $\tau > 0.7$ across all major transitions.

As an additional check on our method, we retroactively identified the nodes that actually changed the state at each major transition, which we term the ``Actual'' node set. We then calculated the Kendall's $\tau$ between $D$ and the average autocorrelation of the Actual node set in each stable range. Note that the size of the Actual node set is different for each major transition. We plot the $\tau$ values for the Actual node set in Fig.~\ref{fig:each}. These $\tau$ values, marked with crosses, have consistently high $\tau$ values, as expected. Notably, the overall performance of the Lower State and High Input node sets across the different major transitions is similar to that of Actual.

\subsection{Reducing the Sampling Cost of the Sentinel Node Sets}
\label{sec:sampling}

We stated in the main text that an advantage of using a sentinel node set---that is to monitor particular $n$ nodes for early warning signals, where $n \ll N$---is reduction in sampling cost. However, in the simulations we have presented in the main text, we determine the $n$ nodes based on the $M$ samples (i.e., time points) of $x_i$ for all the $N$ nodes. This is because we need to rank the $N$ nodes to select the top $n$ nodes. Therefore, strictly speaking, our sentinel node set methods are not efficient in terms of the sampling cost. In this section, we examine a cost-efficient method to only use the state of $x_i$ $\forall i \in \{1, \ldots, N\}$ at the first ten time points to determine the $n$ sentinel nodes. For example, we calculate $R_i = \sum_{j=1}^N a_{ij} \overline{x}_j$, where $\overline{x}_j$ is now the average of $x_j$ over the first ten samples, and use the $n=5$ nodes with the largest $R_i$ as the High Input node set.
Once we have selected the $n$ sentinel nodes in this manner, we use the full $M = 250$ samples, including the first ten samples used for determining the $n$ nodes, to calculate early warning signals.

\begin{figure}
  \includegraphics[width = \textwidth]{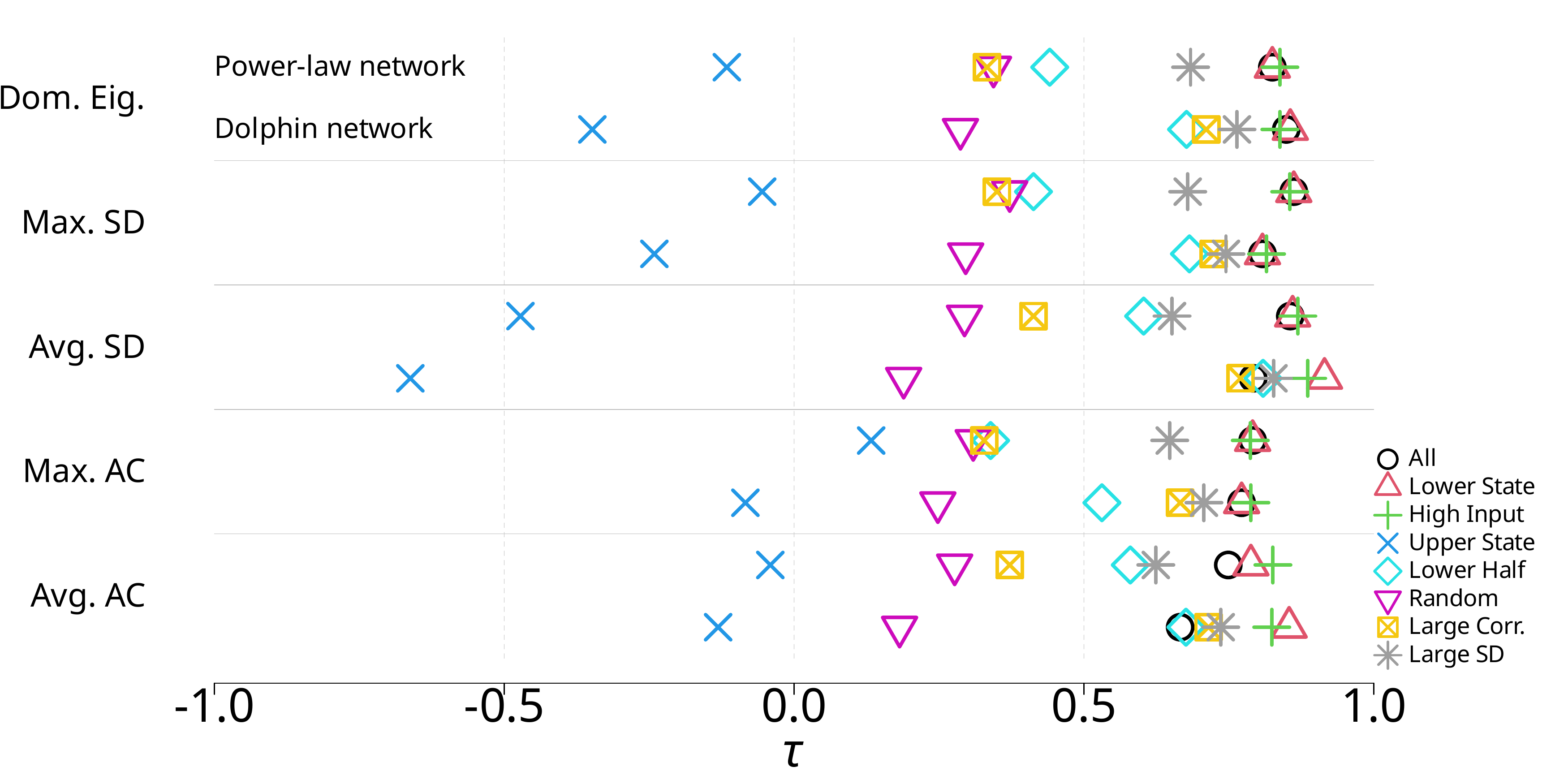}
  \caption{Kendall correlation, $\tau$, between $D$ and each early warning signal for different node sets. Here, we only use the first 10 samples of $x_i$ to determine the sentinel nodes, whereas we use all $M = 250$ samples to calculate early warning signals. See the caption of Fig.~\ref{fig:u2} for the abbreviations.
}
  \label{fig:sampling}
\end{figure}

We find in Fig.~\ref{fig:sampling} that the proposed cost-aware node selection method does not strongly affect the $\tau$ values. An exception is the Large SD node set, the performance of which is notably worse when one uses only 10 samples for determining the sentinel nodes.

\subsection{Anticipating Transitions Without Sufficiently Long Stable Ranges}
\label{sec:nostablerange}

In the analysis in the main text, we identified sufficiently long stable ranges of the bifurcation parameter primarily to assist with Kendall correlation analysis. However, in practice, we do not know whether a sufficiently long stable range precedes a transition to be anticipated. To assess the performance of our method for transitions that are not necessarily preceded by a sufficiently long stable range, we conducted the following analysis for each of the 23 networks with our default settings. For any $D$ value at which any number of nodes transitioned from the lower to the upper state, we identified the set of nodes making the transition, which we denote by $S$. Note that the transition may not be preceded by a sufficiently long stable range. We then identified the node sets with $n=5$ nodes, i.e., High Input, Random, Large Correlation, and Large SD, at the same value of $D$. We say that any of these four node sets is successful at this $D$ value if (a) $|S| < 5$ and $S$ is a subset of the node set, or (b) $|S| \ge 5$ and all nodes in the node set transitioned, where $|S|$ is the number of nodes in $S$. Otherwise, we attribute an error to the node set.

\begin{figure}
  \includegraphics[width = .5\textwidth]{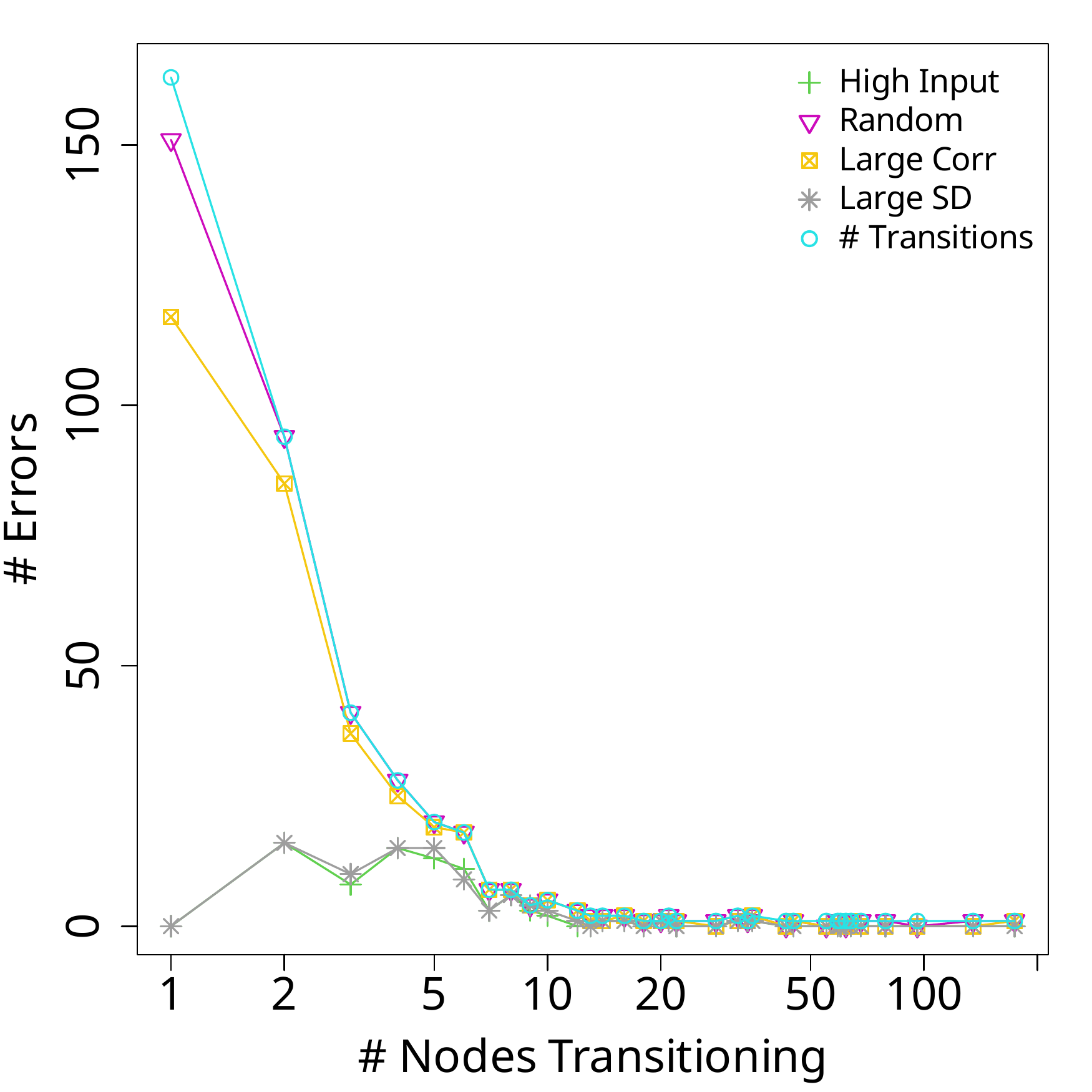}
  \caption{Number of errors in predicting the tipping nodes in any transitions including those not following a sufficiently long stable range of $D$. The plus, triangle, box-times, and star symbols show, for each node set, the average of the number of errors over the transitions from the lower to upper state with the same number of tipping nodes. The circles show the total number of transitions at each size.
  }
  \label{fig:errors}
\end{figure}

We show the number of errors for the different transitions and different node sets for each of the 23 networks in Fig.~\ref{fig:errors}. For example, given a node set, a marker at 1 on the horizontal axis represents the number of transitions involving just one node at which an error occurs. The circles in Fig.~\ref{fig:errors} show the total number of transitions involving each number of nodes.
Figure~\ref{fig:errors} indicates that, when many nodes transition, even a uniformly randomly chosen set of five nodes (i.e., the Random node set) is included in $S$ (see the triangles) simply because $S$ contains many nodes. In contrast, the Random node set makes many errors when few nodes transition. The Random node set provides a baseline. The Large Correlation node set does somewhat better than Random when a single node transitions but performs the same as the Random node set when more nodes transition at the same time. The High Input and Large SD node sets perform the best and retain their advantage for transitions involving over 20 nodes. Thus, the High Input and Large SD node sets, and to a lesser extent the Large Correlation node set, can identify transitions of some nodes even if they do not follow a sufficiently long stable range of the bifurcation parameter.

\subsection{Effectiveness of the Large Correlation Node Set}
\label{sec:largecorr}

\begin{figure}
  \includegraphics[width = .5\textwidth]{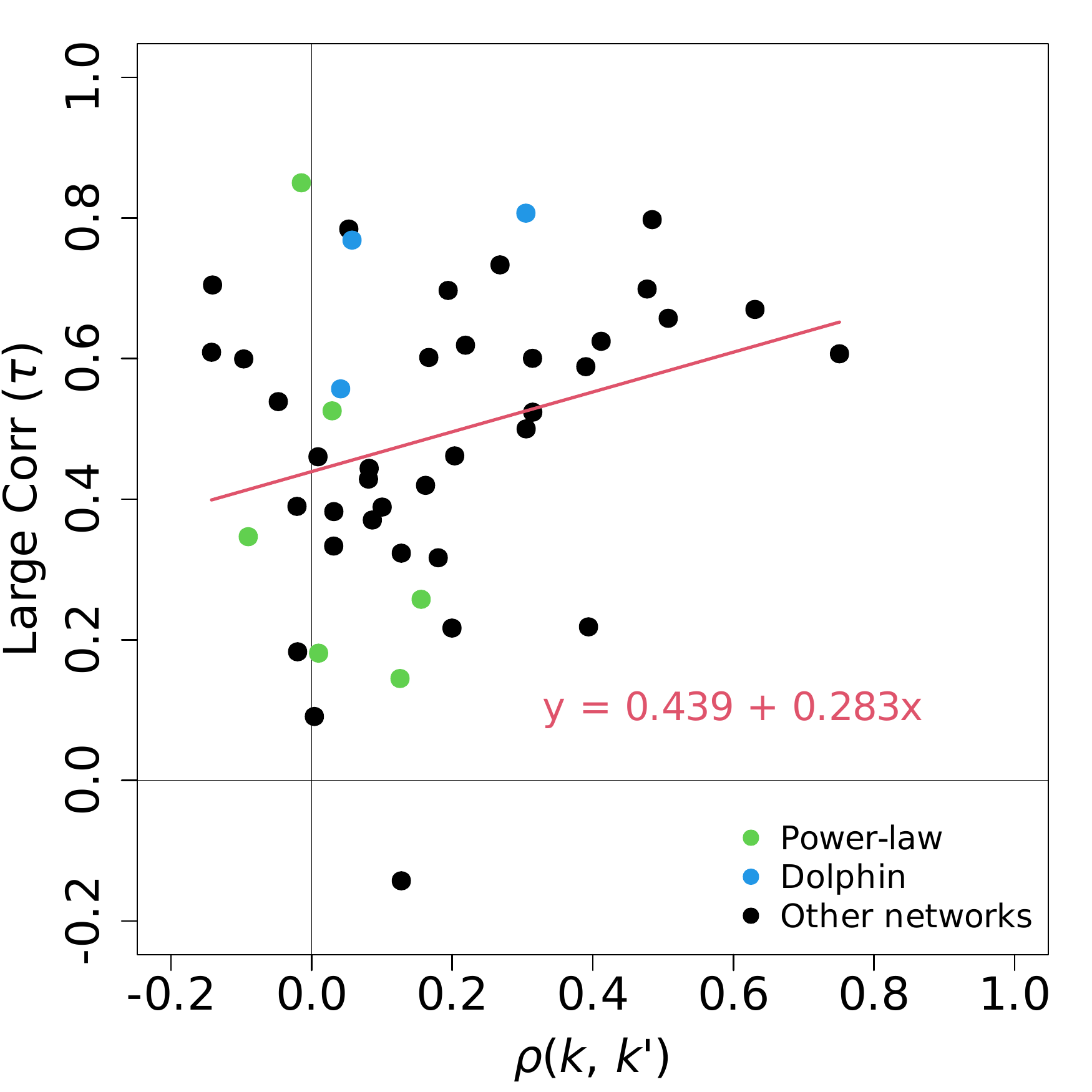}
  \caption{Relationship between the Kendall's $\tau$ for the Large Correlation node set and $\rho(k, k')$. The latter is the Spearman correlation coefficient between the node's degree, $k_i$, and its estimate using the Pearson correlation between the state variables, $k'_i$. A circle represents a major transition in
  one of the 23 networks. Different networks have different numbers of major transitions, contributing different numbers of circles in the figure.
The transitions for the power-law and dolphin networks are marked in green and blue, respectively. The red solid line shows the simple linear relationship between the two variables. The horizontal and vertical solid lines in black show zero correlation values. 
}
  \label{fig:threecorr}
\end{figure}

There is a marked discrepancy in the performance of the Large Correlation node set between the power-law and dolphin networks (see Fig.~\ref{fig:examples-taus} in the main text). This result contrasts with the performance of the High Input node set, which is consistently high for both networks. Because the idea behind Large Correlation is to approximate the presence or absence of an edge between nodes $i$ and $j$, i.e., $A_{ij} \in \{0, 1\}$, by the Pearson correlation between $x_i(t)$ and $x_j(t)$, i.e., $\text{cor}(x_i, x_j)$, we hypothesized that the Large Correlation node set performs better when
$\text{cor}(x_i(t), x_j(t))$ approximates $A_{ij}$ better. To test this hypothesis, we compared
the degree of each $i$th node, denoted by $k_i$, and the degree of the same node estimated by
$k'_i \equiv \sum_{j=1}^N \text{cor}(x_i, x_j)$, $j \neq i$, at each $D$ value at which a major transition occurs.
Specifically, we calculated the Spearman correlation coefficient between $k'_i$ and $k_i$, denoted by $\rho(k, k')$. In Fig.~\ref{fig:threecorr}, we show the relationship between the Kendall's $\tau$ value and $\rho(k, k')$ for the 23 networks. A circle in the figure represents a major transition.  The major transitions in the power-law and dolphin networks are marked in green and blue, respectively.

Figure~\ref{fig:threecorr} shows an apparent positive correlation between $\tau$ and $\rho(k, k')$. The red solid line shows the simple linear regression for the data points; the $p$-value for the coefficient on $\rho(k, k')$ is 0.0047. Therefore, the Large Correlation node set tends to perform better when $k'$ is more similar to $k$, which supports our hypothesis.

\subsection{Multistage Transitions from the Upper State}
\label{sec:multistage}

\begin{figure}
  \centering
  \includegraphics[width = .7\textwidth]{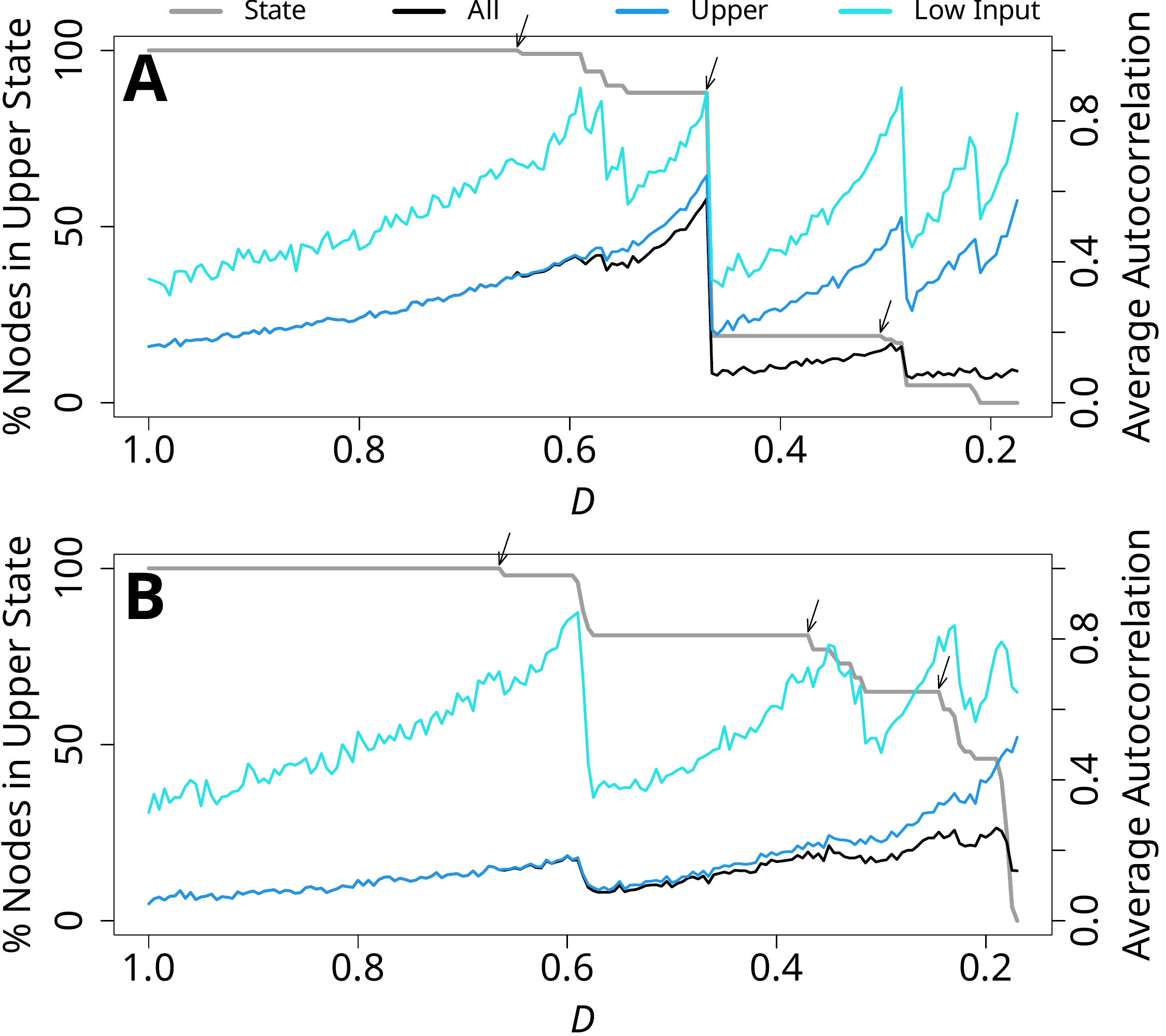}
  \caption{Multistage transitions when the nodes are initially in the upper state. We show the number of nodes in the upper state at equilibrium (gray), and the average lag-1 autocorrelation of $x_{i,t}$ calculated for all nodes (black), the nodes in the upper state (dark blue), and the low-input nodes (light blue). Black arrows mark transitions of nodes at the end of stable ranges. (A) Power-law network. (B) Dolphin network. Note that the horizontal axis is reversed.}
  \label{fig:examples-upper}
\end{figure}

We conducted simulations on the power-law and dolphin networks with all the nodes starting in the upper state, i.e., $x_i = 7~\forall~i$. We set $u = -15$ in Eq.~\eqref{eq:alt} for each value of $D$. We started with $D = 1.0$ and decreased $D$ by 0.005 between each simulation until at least 90\% of nodes were in the lower state at equilibrium. In Fig.~\ref{fig:examples-upper}, we show the proportion of nodes in the upper state and the average autocorrelation calculated for three node sets. As for simulations with nodes beginning in the lower state, we observe multistage transitions for both networks.

\subsection{Statistical Results for the Mixed Linear Effects Model}
\label{sec:lme-table}

We present here the coefficient estimates, standard errors, and $t$ and $p$ values for the linear mixed effects models discussed in the main text. Table \ref{tab:lme} contains the results of model estimation. In the table, the overall model intercept is the reference and corresponds to the ``All'' node set. Additionally, because we use a random effect to capture the effect of different network structures, each network is associated with its own adjustment to the overall intercept. These values form a distribution of random intercepts attributed to network structure. The standard deviation of the distribution of random intercepts for each separate model (that is, for each early warning signal) was 0.024 (dominant eigenvalue), 0.022 (maximum standard deviation), 0.026 (average standard deviation), 0.023 (maximum autocorrelation), and 0.041 (average autocorrelations).

\begin{table}
  \centering
  \begin{tabular}{lcccc}
    \toprule
    Dominant eig. & Est. & SE & $t$ & $p$\\
    \midrule
    Intercept & 0.815 & 0.008 & 106.153 & $< 1.0 \times 10^{-4}$\\
    Lower & 0.003 & 0.002 & 1.959 & 0.0503\\
    Sentinel & 0.015 & 0.002 & 8.712 & $< 1.0 \times 10^{-4}$\\
    \midrule
    Maximum SD & Est. & SE & $t$ & $p$\\
    Intercept & 0.824 & 0.007 & 117.584 & $< 1.0 \times 10^{-4}$\\
    Lower & 0.002 & 0.002 & 1.362 & 0.1734\\
    Sentinel & 0.004 & 0.002 & 2.244 & 0.025\\
    \midrule
    Average SD & Est. & SE & $t$ & $p$\\
    \midrule
    Intercept & 0.792 & 0.009 & 91.429 & $< 1.0 \times 10^{-4}$\\
    Lower & 0.091 & 0.003 & 27.827 & $< 1.0 \times 10^{-4}$\\
    Sentinel & 0.081 & 0.003 & 24.777 & $< 1.0 \times 10^{-4}$\\
    \midrule
    Maximum AC & Est. & SE & $t$ & $p$\\
    \midrule
    Intercept & 0.751 & 0.008 & 99.758 & $< 1.0 \times 10^{-4}$\\
    Lower & 0.002 & 0.002 & 1.059 & 0.2896\\
    Sentinel & 0.015 & 0.002 & 7.708 & $< 1.0 \times 10^{-4}$\\
    \midrule
    Average AC & Est. & SE & $t$ & $p$\\
    \midrule
    Intercept & 0.667 & 0.013 & 50.295 & $< 1.0 \times 10^{-4}$\\
    Lower & 0.122 & 0.003 & 40.354 & $< 1.0 \times 10^{-4}$\\
    Sentinel & 0.145 & 0.003 & 48.093 & $< 1.0 \times 10^{-4}$\\
    \bottomrule
  \end{tabular}
  \caption{Linear mixed effects models explaining the Kendall's $\tau$ over 50 simulation runs and 10 networks showing multistage transitions for three node sets, i.e., All, Lower State, and High Input. The rows labeled Lower State and High Input represent the effect of these node sets relative to All as the reference. Est., SE, and $t$ refer to the estimated coefficient, its standard deviation, and the $t$ statistic, respectively.}
  \label{tab:lme}
\end{table}

\subsection{Parameter Variation}
\label{sec:param-var}

\begin{figure}
  \includegraphics[width = \textwidth]{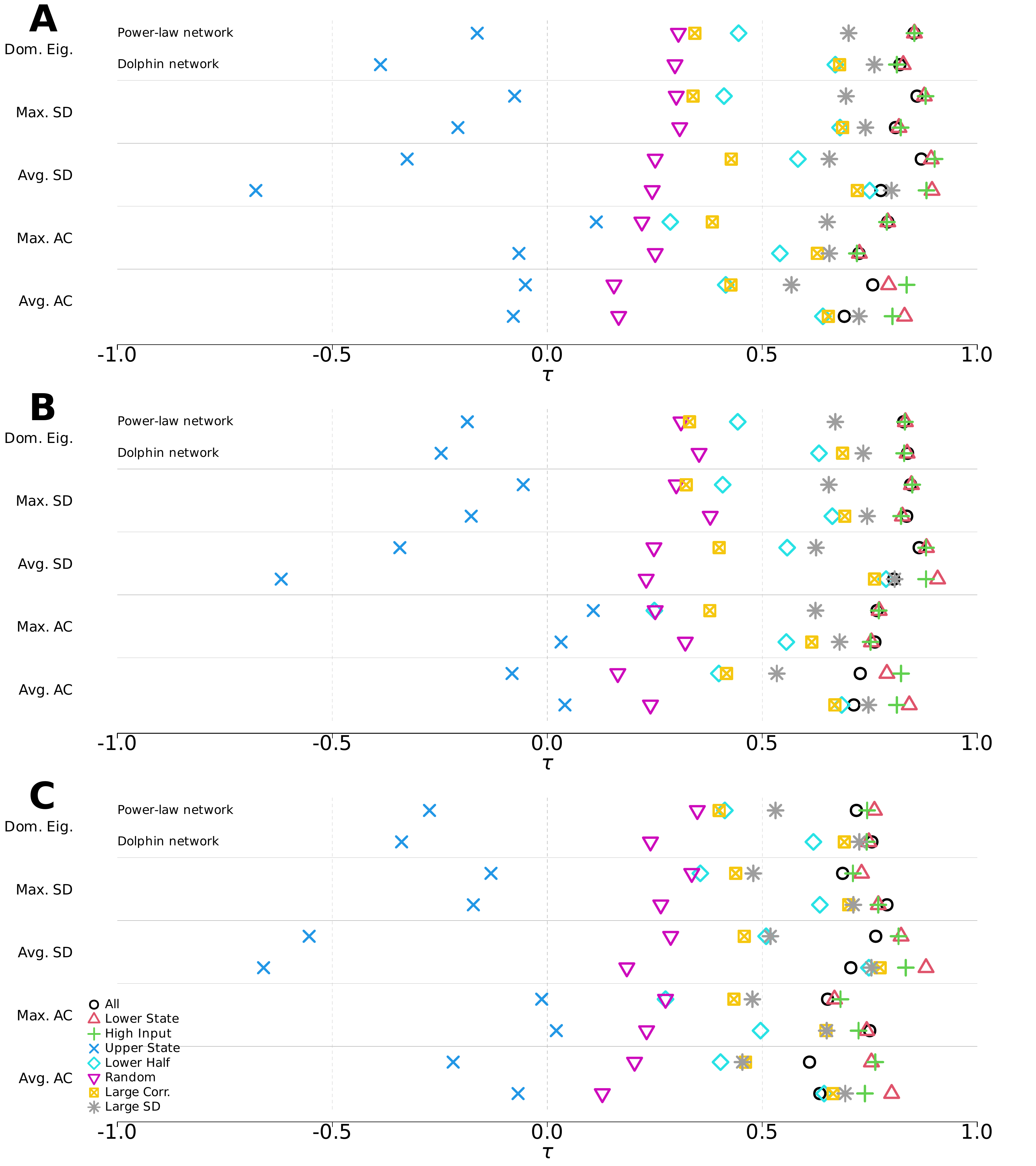}
  \caption{Kendall's $\tau$ when we vary $s$. (A) $s = 0.01$, (B) $s = 0.1$, and (C) $s = 0.5$. Early warning signals are indicated on the vertical axis (Dom. Eig.: dominant eigenvalue, Max. SD: maximum standard deviation, Avg. SD: average standard deviation, Max. AC: maximum autocorrelation, Avg. AC: average autocorrelation). Large Corr. in the legend abbreviates the Large Correlation node set.}
  \label{fig:var-s}
\end{figure}

\begin{figure}
  \includegraphics[width = \textwidth]{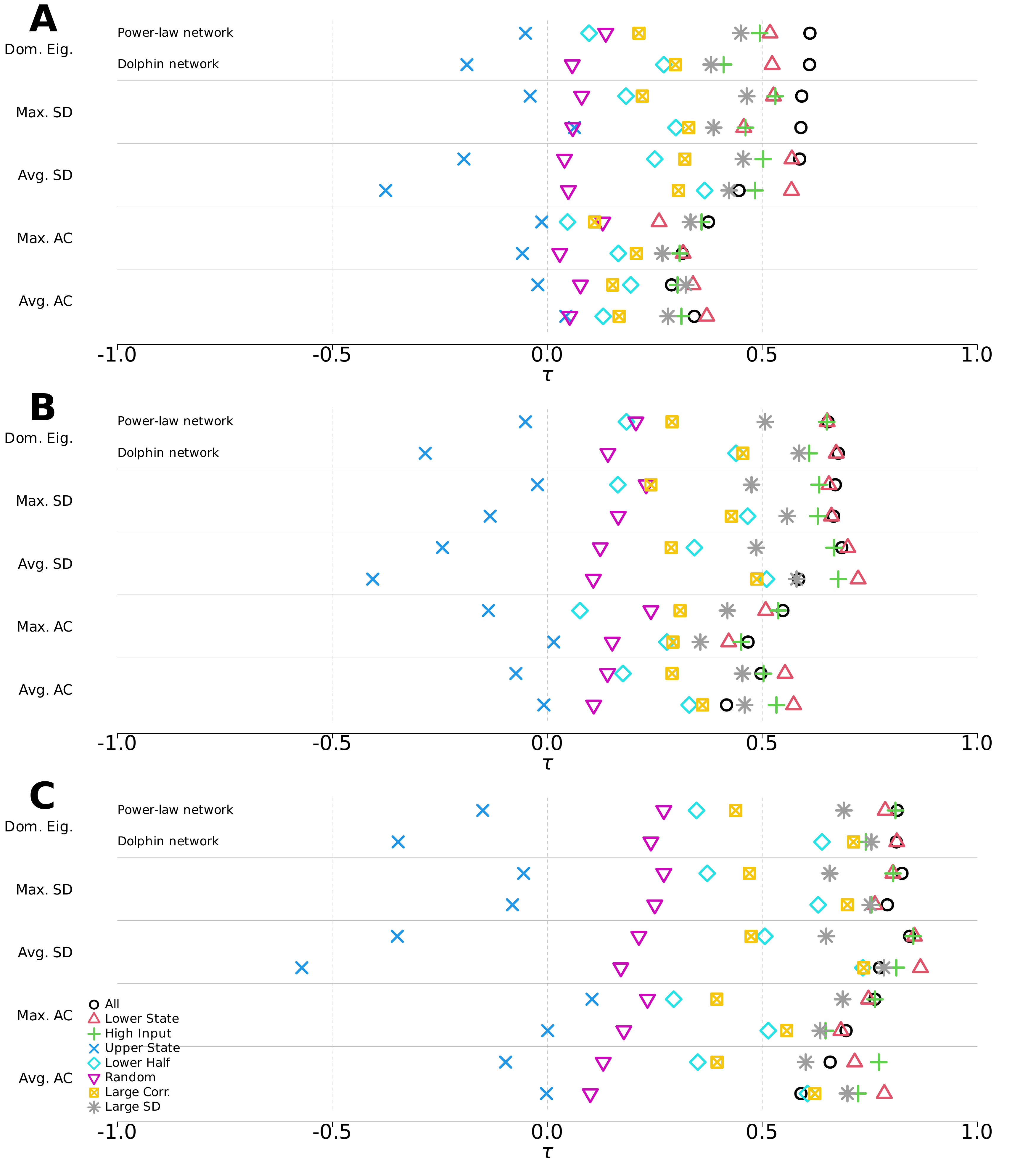}
  \caption{Kendall's $\tau$ when we vary $M$. (A) $M = 25$, (B) $M = 50$, and (C) $M = 150$.}
  \label{fig:var-M}
\end{figure}

\begin{figure}
  \includegraphics[width = \textwidth]{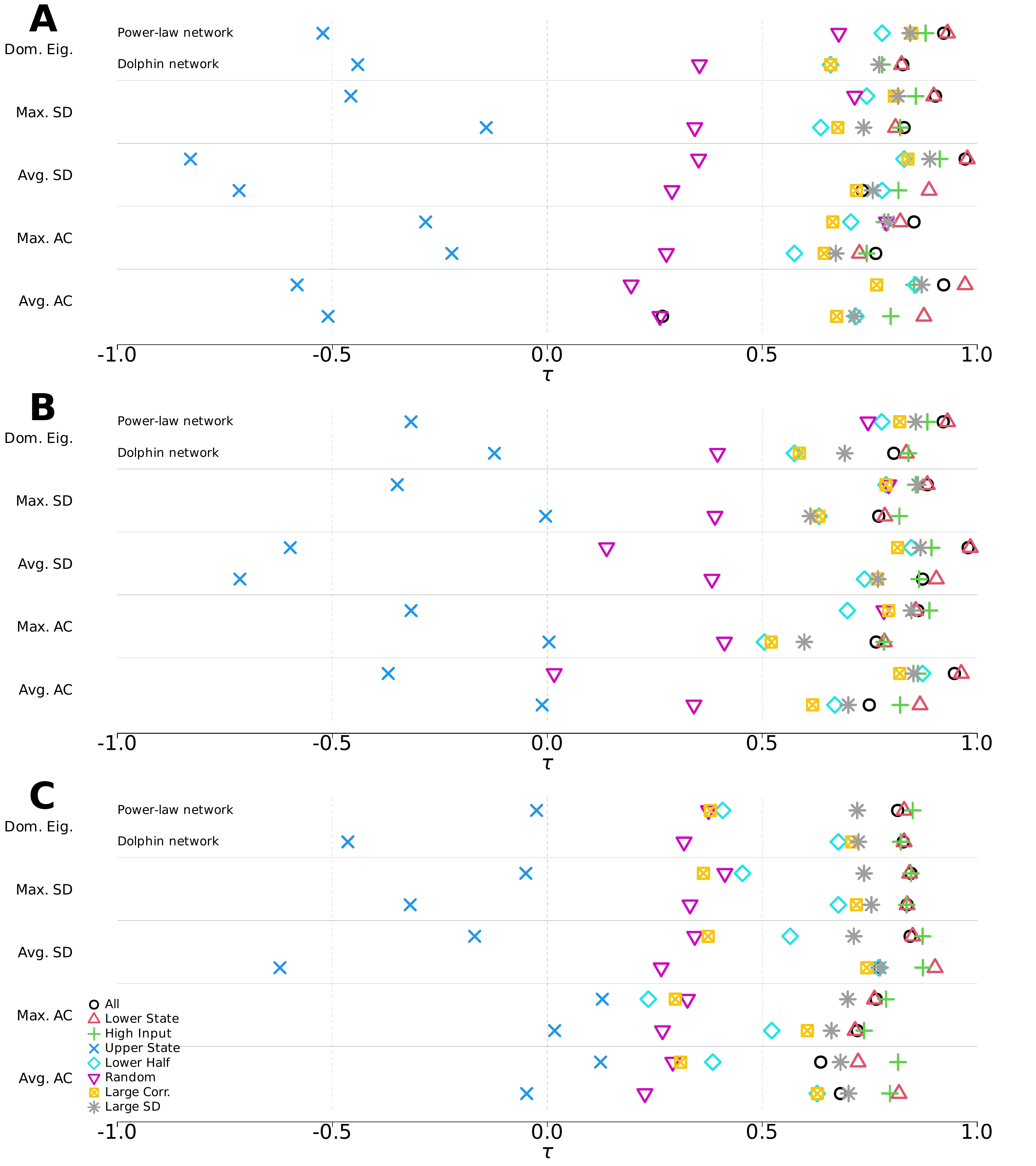}
  \caption{Kendall's $\tau$ when we vary $(r_1, r_2, r_3)$. (A) $(r_1, r_2, r_3) = (1, 3, 5)$, (B) $(r_1, r_2, r_3) = (1, 2.5, 7)$, and (C) $(r_1, r_2, r_3) = (1, 5.5, 7)$.}
  \label{fig:var-r}
\end{figure}

\begin{figure}
  \includegraphics[width = \textwidth]{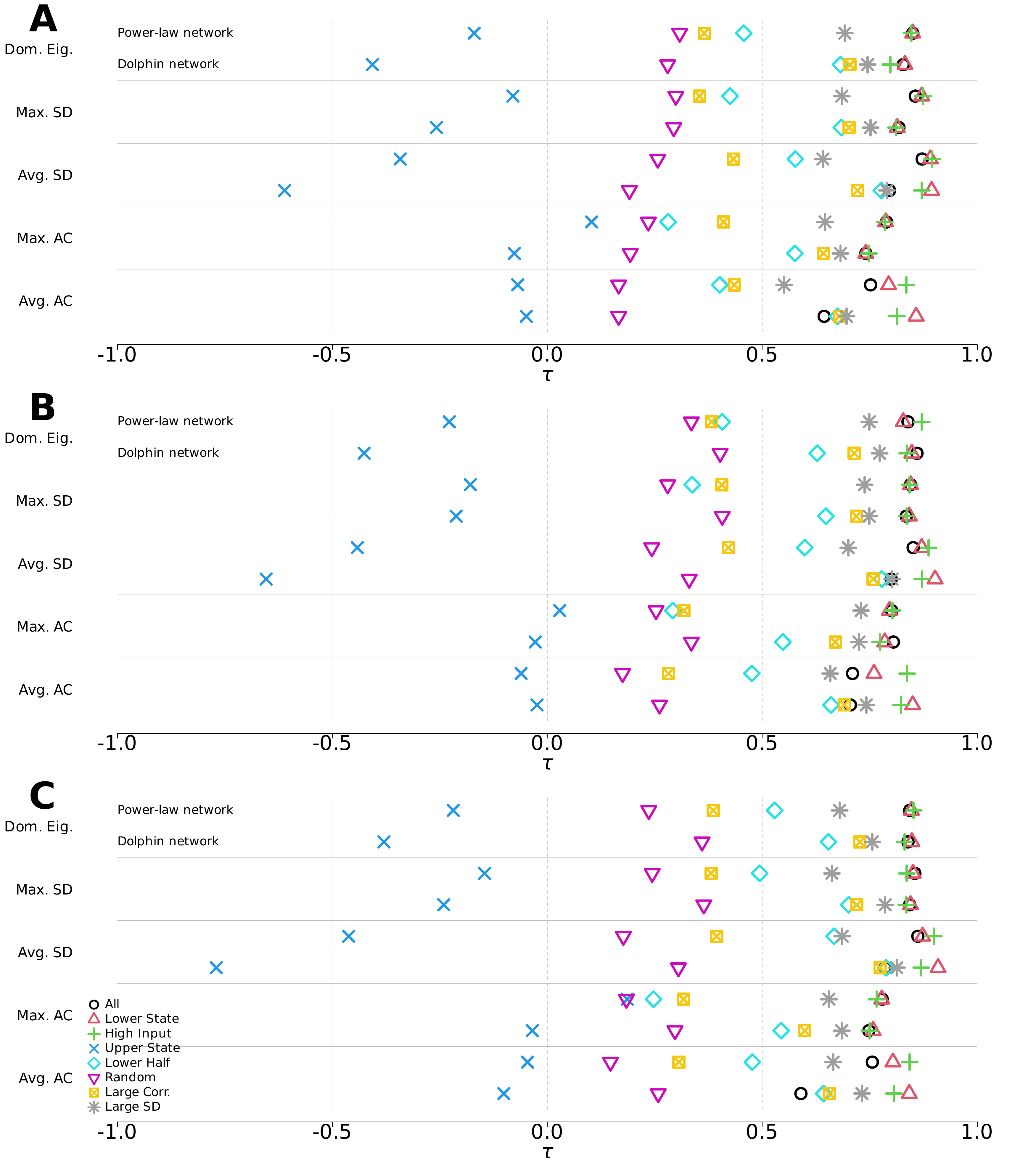}
  \caption{Kendall's $\tau$ when we vary the length of the simulation. The duration of the simulations $T$ prior to early warning signal sampling is (A) 25 TU, (B) 75 TU, and (C) 100 TU.}
  \label{fig:var-TU}
\end{figure}

We conducted additional simulations on the power-law and dolphin networks to investigate the effects of model parameters on the results presented in the main text. We varied noise intensity ($s \in \{0.01, 0.1, 0.5\}$), the number of samples taken from $x_i(t)$ ($M \in \{25, 50, 150\}$), the double-well system's parameters ($(r_1, r_2, r_3) \in \{(1, 3, 5),~(1, 2.5, 7),\allowbreak~(1, 5.5, 7)\}$), and the duration $T$ of the simulation before we sampled $x_i(t)$ ($T \in \{25, 75, 100\}$). Except for the parameter that we varied, we kept all the parameter values to be the same as those used in the main text. 
As in Sect. \ref{sec:sampling}, we assessed node state using only the first 10 time points. We initialized the nodes to be in their lower state in each simulation.
We show Kendall's $\tau$ values for each set of simulations in Figs.~\ref{fig:var-s}--\ref{fig:var-TU}.

\end{document}